\pdfoutput=1
\documentclass[twocolumn,prd,preprintnumbers,amsmath,amssymb,superscriptaddress, nofootinbib]{revtex4-1}
\usepackage{natbib}
\usepackage{graphicx}
\usepackage{amsmath}
\usepackage{amsfonts}   
\usepackage{amssymb}  
\usepackage{mathrsfs}
\usepackage{epsfig,bbm}
\usepackage{color}
\usepackage{multirow}

\newcommand\fverb{\setbox\fverbbox=\hbox\bgroup\verb}
\newcommand\fverbdo{\egroup\medskip\noindent%
\fbox{\unhbox\fverbbox}\ }
\newcommand\fverbit{\egroup\item[\fbox{\unhbox\fverbbox}]}
\newbox\fverbbox

\newcommand{\data}{d}

\newcommand{\params}{\theta}

\newcommand{\like}{{\mathcal L}}

\newcommand{\be}{\begin{equation}}
\newcommand{\ee}{\end{equation}}
\newcommand{\beq}{\begin{equation}}
\newcommand{\eeq}{\end{equation}}
\newcommand{\bea}{\begin{eqnarray}}
\newcommand{\eea}{\end{eqnarray}}
\newcommand{\gsim}{\lower.7ex\hbox{$\;\stackrel{\textstyle>}{\sim}\;$}}
\newcommand{\lsim}{\lower.7ex\hbox{$\;\stackrel{\textstyle<}{\sim}\;$}}

\begin{document}

\title{Fundamental statistical limitations of future dark matter direct detection experiments} 

\author{Charlotte Strege}
\affiliation{Astrophysics Group, Imperial College London,
  Blackett Laboratory, Prince Consort Road, London SW7 2AZ, UK} 
\author{Roberto Trotta}
\affiliation{Astrophysics Group, Imperial College London,
  Blackett Laboratory, Prince Consort Road, London SW7 2AZ, UK} 
  \affiliation{Kavli Institute for Theoretical Physics, University of California, Santa Barbara, CA 93106-4030, USA}
\author{Gianfranco Bertone}
\affiliation{Institute for Theoretical Physics, University of Amsterdam, Science Park 904, Postbus 94485, 1090 GL Amsterdam, The Netherlands} 
\author{Annika H. G. Peter}
\affiliation{Department of Physics and Astronomy, University of California, Irvine, California 92697-4575, USA}
\author{Pat Scott}
\affiliation{Department of Physics, McGill University, 3600 rue University, Montr\'eal, QC, H3A 2T8, Canada}

\date{\today}

\begin{abstract}
\noindent 

We discuss irreducible statistical limitations of future ton-scale dark matter direct detection experiments.  We focus in particular on the coverage of confidence intervals, which quantifies the reliability of the statistical method used to reconstruct the dark matter parameters, and the bias of the reconstructed parameters. We study 36 benchmark dark matter models within the reach of upcoming ton-scale experiments. We find that approximate confidence intervals from a profile-likelihood analysis exactly cover or over-cover the true values of the WIMP parameters, and hence are conservative. We evaluate the probability that unavoidable statistical fluctuations in the data might lead to a biased reconstruction of the dark matter parameters, or large uncertainties on the reconstructed parameter values.  We show that this probability can be surprisingly large, even for benchmark models leading to a large event rate of order a hundred counts. We find that combining data sets from two different targets leads to improved coverage properties, as well as a substantial reduction of statistical bias and uncertainty on the dark matter parameters.

\end{abstract}

\pacs{Suggested PACS number}

\maketitle


\section{Introduction}
\label{sec:intro}

Among the large number of possible dark matter candidates \cite{Bertone:2010zz,Bergstrom:2000pn,Munoz:2003gx,Bertone:2004pz}, weakly-interacting massive particles (WIMP) \cite{steigman1985} are by far the most widely studied.  WIMPs naturally arise from popular extensions of the Standard Model of particle physics (e.g., the lightest neutralino in supersymmetry \cite{griest1988,jungman1996} and the ${B}^1$ in theories with universal extra dimensions \cite{appelquist2001,servant2002,cheng2002}), and they naturally achieve the appropriate cosmological relic density through thermal freeze-out in the early Universe. 

Several experiments are currently searching for these particles by looking for signals of WIMPs scattering on atomic nuclei in large underground detectors, and many others are planned for the next decade (see e.g. Ref.~\cite{Bertone:2010zz} and the discussion in Ref.~\cite{Pato:2010zk}). Although the DAMA/LIBRA \cite{Bernabei:2010mq} and CoGeNT \cite{Aalseth:2010vx} collaborations have reported a modulation of the measured event rate that has been tentatively interpreted in terms of WIMPs (e.g.~\cite{Arina:2011zh}), and the CRESST-II collaboration has found a large excess of events in the acceptance region where a WIMP signal would be expected \cite{Angloher:2011uu}, these results can hardly be reconciled with null searches from experiments such as XENON100 \cite{Aprile:2010um,Aprile:2011hi}, CDMS-II \cite{Ahmed:2009zw,Ahmed:2010wy}, EDELWEISS-II \cite{Armengaud:2011cy} and ZEPLIN-III \cite{Akimov:2011tj}. The controversy will hopefully be resolved by next-generation direct detection experiments, where larger rates and better statistics could lead to an incontrovertible discovery of dark matter. 

If a WIMP-nucleon scattering signal is detected, the event rate and the shape of the measured spectrum of recoil energies can be used to determine the properties of the dark-matter particle, most importantly its mass and scattering cross-section. The constraining power of present and upcoming experiments has been thoroughly discussed in the literature \cite{Green:2008rd,strigari2009,Pato:2010zk,akrami2011,peter2011}. Here, we present {\it irreducible statistical limitations} of future dark matter direct detection experiments.

We focus on two different issues: first, we explore the concept of {\it coverage} of confidence intervals, which quantifies the reliability of the statistical method adopted to reconstruct the WIMP parameters.  We investigate the coverage of one-dimensional confidence intervals, constructed using an approximate method that relies on the assumption that profile likelihood ratios are chi-square distributed, based on Wilks' theorem \cite{Wilks1938}. This approximate method of constructing confidence intervals is commonly used for frequentist data analysis in the literature in lieu of more complex methods (e.g. Feldman and Cousins \cite{PhysRevD573873}), which provide exact coverage by construction.  The coverage of parameter reconstructions has been previously discussed in the context of direct detection \cite{Akrami:2010cz} and collider identification \cite{Bridges:2010de} of supersymmetric models.

Second, we consider how well one can expect to reconstruct the WIMP properties from future direct-detection data, given the statistical fluctuations that will inevitably impact the observed energy spectrum.  We perform parameter reconstructions on thousands of simulated data sets to estimate the \emph{average} uncertainty and bias in the reconstructions of several different WIMP benchmark models. Additionally, we provide an estimate of the number of \emph{outliers} in the parameter reconstructions. We show that for several different benchmark models that lead to small average uncertainties in the parameter reconstruction, a non-negligible percentage of all reconstructions results in a much larger uncertainty, as a result of statistical fluctuations that impact on each individual data set. Considering the number of outliers for different WIMP benchmark models is of crucial importance, since in practice there will be a unique realisation of each experiment, and the constraints derived from a particular realisation can be very different from the outcome for the ``average experiment", as illustrated below. Finally, we investigate how the average uncertainty in the WIMP mass can be decreased by increasing the exposure of direct detection experiments, for several different benchmark points in WIMP parameter space.

The complementarity between direct detection experiments using different target materials, and the possibility of obtaining tighter constraints on the WIMP parameters when combining data from more than one experiment, have recently been emphasized in Ref.~\cite{akrami2011,peter2010,Pato:2010zk,McDermott:2011hx}. Here we compare the coverage, uncertainty and bias of reconstructed parameters for various benchmark points, based either on mock data sets from a single xenon experiment, or a combined analysis of mock data from a xenon experiment and a germanium experiment.

Throughout our analysis we assume that the background event rate is negligible, and ignore uncertainties in the nuclear physics of elastic scattering and the local WIMP distribution function.  We expect that the coverage, precision and bias of our reconstructions will degrade if the backgrounds are non-negligible and astrophysical uncertainties are fully taken into account.  Given this optimistic set-up, we present here a set of \emph{irreducible limitations} on WIMP parameter reconstruction from future direct-detection experiments, arising from fundamental statistical fluctuations driven by the Poisson nature of the event rate.

The paper is organized as follows: in Sec. \ref{DD} we introduce the formalism of direct dark matter detection and discuss the expected performance of upcoming experiments. In Sec. \ref{statsection} we present our parameter reconstruction method and introduce the statistical quantities we use to quantify the performance of our reconstruction procedure. We present our results in Sec. \ref{results} and our conclusions in Sec. \ref{sec:summary}.
 

\section{Direct dark matter detection}\label{DD}

\subsection{Theoretical formalism}
\label{DDtheory}

\par Dark matter direct detection experiments aim to detect signals of WIMPs scattering on target nuclei. The nuclear recoil spectrum for a WIMP of mass $m_{\chi}$ and a target nucleus of mass $m_N$ has the form   
\begin{equation}\label{recoil1}
 \frac{dR}{dE_R}(E_R)=\frac{\rho_0}{m_{\chi}m_{N}} \, \int_{v>v_\text{min}}
 { d^3\vec{v} \,  \frac{d\sigma}{dE_R} v f\left(\vec{v}+\vec{v_E}\right)} \quad. 
\end{equation}
Here $dR/dE_R$ has units of events per unit energy per unit time per unit target material mass, $\rho_0$ is the local dark matter density, $\sigma$ is the WIMP-nucleus scattering cross-section and $E_R$ is the WIMP-induced recoil energy of the nucleus. Neglecting gravitational focusing of WIMPs as they flow into the potential well of the Solar System, $f(\vec{u})$ is the normalized local WIMP velocity distribution function in the rest frame of the Galaxy, $\vec{v_E}$ is the Earth's velocity in this frame and $\vec{v}$ is the velocity of the WIMPs in the rest frame of the Earth (which is also the WIMP-nucleon relative velocity, as to a good approximation the nucleons are at rest in the Earth frame). In this paper we focus on elastic WIMP-nucleus interactions. For elastic scattering the minimum velocity $v_\text{min}$ required for a WIMP of mass $m_{\chi}$ to be able to induce a nuclear recoil of energy $E_R$ is 
\beq
v_\text{min} = \sqrt{\frac{m_N E_R}{2 \mu^2_N}} \quad,
\eeq
where $\mu_N = m_{\chi}m_N/(m_{\chi} + m_N)$ is the WIMP-nucleus reduced mass. 

The differential scattering cross-section $d\sigma/dE_R$ includes different types of WIMP-nucleus interactions. We will assume that all events result from spin-independent WIMP-nucleus scattering and neglect all other types of interactions. In this case the differential scattering cross-section is given by
\begin{equation}
\frac{d\sigma}{dE_R}=\frac{m_N}{2 v^2 \mu_N^2} \sigma^{SI}_N \mathcal{F}^2(E_R) \quad,
\end{equation}
where $\mathcal{F}(E_R)$ is the spin-independent nuclear form factor, which accounts for the finite extent and composite nature of the atomic nucleus, and $\sigma^{SI}_N$ is the spin-independent (SI) zero-momentum WIMP-nucleus cross-section. This cross-section can be written in terms of the mass number of the nucleon $A$, its atomic number $Z$, the WIMP-proton coupling $f_p$, and the WIMP-neutron coupling $f_n$.
\beq
\label{cross-sec}
\sigma^{SI}_N = \frac{4}{\pi} \mu_N^2 (Z f_p + (A - Z) f_n)^2 \quad.
\eeq
In the following we will assume that the WIMP-proton and WIMP-neutron couplings are very similar $f_p \sim f_n$ (as appropriate in most supersymmetric setups \cite{Ellis:2008hf}, but see also Refs.\cite{Chang:2010yk,Feng:2011vu,Kang:2010mh,Pato:2011de} for alternative scenarios), so that the WIMP-nucleus cross-section simplifies to $\sigma^{SI}_N = 4 \mu_N^2 A^2 f_p^2 /\pi$. In analogy to this expression we define the WIMP-proton cross-section $\sigma_p^{SI} = 4 \mu_p^2 f_p^2 / \pi$, with $\mu_p = m_{\chi} m_p/ (m_{\chi} + m_p)$ the WIMP-proton reduced mass. The differential scattering cross-section can then be rewritten as
\begin{equation} \label{sigma}
\frac{d\sigma}{dE_R} = \frac{m_N}{2 v^2 \mu_p^2} A^2 \sigma^{SI}_p \mathcal{F}^2(E_R) \quad.
\end{equation}
In this analysis we use the Helm form factor \cite{1996APh.....6...87L}
\beq
\mathcal{F}(E_R)=3\frac{\textrm{sin}(q r)-(q r)\textrm{cos}(q r)}{(q r)^3} e^{-(q s)^2/2} \quad,
\eeq
where $q = \sqrt{2 m_N E_R}$ is the momentum transferred in the recoil, $s = 0.9$ fm, $r = \sqrt{c^2 + 7\pi^2a^2/3 - 5s^2}$, $a = 0.52$ fm and $c = (1/23A^{1/3} - 0.6)$ fm.
Using Eq.~\eqref{sigma} the nuclear recoil spectrum can be rewritten as
\begin{equation} 
\label{recoil2}
\frac{dR}{dE_R} (E_R) =\frac{\rho_0\sigma^{SI}_p A^2 \mathcal{F}^2(E_R)}{2 \mu_p^2 m_{\chi}} \int_{v>v_\text{min}} d^3\vec{v} \, \frac{f\left(\vec{v}+\vec{v_E}\right)}{v} \quad.
\end{equation}
The quantities of interest are the WIMP mass $m_{\chi}$ and the spin-independent WIMP-proton cross-section $\sigma^{SI}_p$. The choice of target material enters the analysis via the mass number $A$ and the form factor $\mathcal{F} (E_R)$, and through $v_\text{min}$. Note for $m_\chi \gg m_N$, $v_\text{min} \rightarrow \sqrt{E_R/2m_N}$, and hence the recoil spectrum depends on $m_\chi$ and $\sigma_p^{SI}$ only via the degenerate combination $\sigma_p^{SI}/(\mu_p^2 m_\chi)$, which has a strong impact on the performance of the reconstruction of the WIMP properties, as we will see in the following sections.  

The third component that enters the recoil rate is the local astrophysical DM distribution, most importantly the local density $\rho_0$ and the WIMP velocity distribution $f(\vec{u})$. In this analysis we will model local astrophysics using the standard halo model.   This model consists of an isothermal, spherically symmetric galactic WIMP distribution.  In this model, WIMP velocities follow a non-rotating isotropic Maxwellian distribution in a Galactocentric frame with a one-dimensional velocity dispersion $v_0/\sqrt{2}$, where $v_0$ is the speed of the Local Standard of Rest. WIMPs traveling at very high velocities will escape the gravitational attraction of the galaxy and will therefore not be present in the halo. This is taken into account by truncating the velocity distribution at some escape velocity $v_{esc}$, leading to a WIMP velocity distribution function
\beq \label{Maxw}
f(\vec{v}+\vec{v_E})=\left\{
\begin{array}  {c l}
\frac{N^{-1}}{v_0^3 \pi^{3/2}} e^{-(\vec{v}+\vec{v_E})^2/v_0^2}, \ &\textnormal{for} \ |\vec{v}+\vec{v_E}| < v_{esc} \\
0 &\textnormal{otherwise} \quad,
\end{array}\right.
\eeq   
with $N = \textnormal{erf}(v_{esc}/v_0) - 2\pi^{-1/2}(v_{esc}/v_0)e^{-(v_{esc}/v_0)^2}$ a normalization factor which ensures that $\int d^3 \vec{u} \ f(\vec{u}) = 1$. The velocity of the Earth with respect to the rest frame of the galaxy is given by the sum of the local circular velocity $\vec{v_0}$, the Sun's peculiar velocity $\vec{v_{\textnormal{pec}}}$ and the Earth's velocity relative to the Sun $\vec{v_{\textnormal{orb}}}$
\be \label{earth vel}
\vec{v_E} = \vec{v_0} + \vec{v_{\textnormal{pec}}} + \vec{v_{\textnormal{orb}}} \quad.
\ee
The contribution of both $|\vec{v_{\textnormal{pec}}}| \sim 10$ km/s and $\vec{v_{\textnormal{orb}}} \sim 30$ km/s to $\vec{v_E}$ is small compared to the contribution of $\vec{v_0} \sim 200 - 300$ km/s. As we consider neither directional signatures nor the annual modulation of the nuclear recoil spectrum in this study, the latter two terms in Eq.~\eqref{earth vel} can be neglected and $\vec{v_E} \simeq \vec{v_0}$.

It is well known that there is a sizeable uncertainty on the astrophysical parameters $\rho_0$,$v_0$,$v_{esc}$ and $f(\vec{u})$. Additionally, the standard halo model can only be considered a first approximation to a much more complicated halo profile \cite{read2009,vogelsberger2009,kuhlen2010,lisanti2011}. In order to achieve a correct reconstruction of the WIMP parameters from experiment, it is of vital importance to take into account these uncertainties \cite{strigari2009,akrami2011,peter2011}. The aim of this paper is to investigate the coverage properties and the quality of the reconstruction for different WIMP benchmark models and identify any unavoidable statistical effects. In order to do so we will assume an ideal case, fixing all of the astrophysical parameters to their fiducial values and neglecting their uncertainties. The fiducial values we use are $\rho_0 = 0.4 \ \textnormal{GeV/cm}^3$, $v_0 = 230$ km/s and $v_{esc} = 544$ km/s.  We will investigate coverage properties of a more general framework that includes astrophysical uncertainties in the WIMP distribution function in a future work. 

The total number of recoil events $N_R$ can be found by weighting the nuclear recoil rate in Eq.~\eqref{recoil2} by the event acceptance $\epsilon(E_R)$, and integrating from some threshold energy $E_{thr}$ to some maximum energy $E_\text{max}$.  Assuming that the acceptance is not energy-dependent, $\epsilon(E_R)$ simply falls out of the integral, and becomes a mean effective exposure $\epsilon_\text{eff}$ (which is the product of the detector mass and exposure time).  $N_R$ is then given by  
\begin{equation}
\label{nocounts}
N_R = \epsilon_\text{eff} \int_{E_{thr}}^{E_\text{max}}{ dE_R \, \frac{dR}{dE_R} } \quad.
\end{equation}

For our coverage study, we select a number of WIMP benchmark models, with benchmark mass and cross-section ranges $m_{\chi} = [25,250]$ GeV and $\sigma_p^{SI} = [10^{-8},10^{-10}]$ pb. For each benchmark point the analysis is based on $10^3$ mock data sets.

\subsection{Future direct detection experiments} \label{exp}

In order to assess the performance of the reconstruction of WIMP properties from next-generation direct detection data, we will use ton-scale, low-background versions of two current detectors.  We will systematically investigate the constraints that data sets from these experiments can place on the WIMP properties for different benchmark models. 

The most stringent constraints on WIMP properties are currently provided by the XENON100 collaboration \cite{Aprile:2011hi}. The recently published $90\%$ C.L. exclusion curve has a minimum cross-section of $\sigma_p^{SI} = 7.0 \times 10^{-9}$ pb at a WIMP mass $m_{\chi} = 50$ GeV \cite{Aprile:2011hi}. These constraints will be improved further once data from the proposed XENON1T experiment becomes available in 2015 \cite{aprile2011}. Additionally, the DARWIN project\footnote{http://darwin.physik.uzh.ch} is working towards a multi-ton scale noble liquid experiment which is expected to start running in 2017 and will probe spin-independent cross-sections down to $10^{-12}$ pb. \cite{Baudis:2012bc} A second promising WIMP detection strategy is based on cryogenic detectors operating at very low temperatures, most notably the current CDMS-II germanium experiment \cite{Ahmed:2009zw}. The SuperCDMS and GEODM cryogenic germanium experiments aim to upgrade this experiment to the ton scale within the next decade \cite{figueroa2010}. A second planned experiment using cryogenic detectors operating at mK temperatures is EURECA\footnote{http://www.eureca.ox.ac.uk}. This experiment is pushing for a target mass of 1 ton and will probe cross-sections down to $10^{-10}$ pb.

In this study we will use a ton-scale experiment with a liquid natural Xe target with average atomic mass 131 g/mol, and a ton-scale Ge experiment with atomic mass 73 g/mol. The characteristics of these detectors are chosen to reflect projects that can realistically be built within the next 5 - 10 years; they are given in Table \ref{experiments}.  Although large liquid argon experiments are also currently under construction, we choose not to include simulated argon data in this study, because previous studies have shown that germanium and xenon provide tighter constraints on the WIMP parameters and halo velocity distribution \cite{Pato:2010zk}.

\begin{table*}[htp]
\centering
\fontsize{9}{9}\selectfont
\begin{tabular}{|c|c|c|c|c|c|}
\hline
\hline
 Target &  $E_{thr}$ [keV] & $\epsilon$ [ton$\times$year] & $A_{NR}$ & $\epsilon_\text{eff}$ [ton$\times$yr] & \# Background events \\
\hline
Xe & 10.0 & 5.00 & 0.5 & 2.00 & $<1$ \\
Ge & 10.0 & 3.00 & 0.9 & 2.16 & $<1$ \\
\hline
\end{tabular}
\caption{\fontsize{9}{9}\selectfont Primary characteristics of future ton-scale dark matter direct detection experiments using xenon and germanium as target materials. For further details see section \ref{exp}.} \label{experiments}
\end{table*}

For both the xenon and the germanium experiment we assume a threshold energy of $E_{thr} = 10$ keV and only consider recoil energies below $100$ keV.  This is a reasonable cut-off, given the exponential decay of the WIMP-nucleus recoil spectrum with energy.  Studies have shown that resolving the exponential decay at high energies is important for improving parameter reconstruction \cite{peter2011}.  For both experiments we assume a total cut efficiency of $\eta_\text{cut} = 80\%$. Following Ref.~\cite{Pato:2010zk}, for the Xe experiment we take a fiducial detector mass of 5 tons and one year of operation. We assume that a percentage $A_{NR} = 50\%$ of all nuclear recoils in the fiducial region are accepted, so that, after inclusion of the overall cut efficiency, the effective exposure is $\epsilon_\text{eff} = 2.00$ ton$\times$year. For the germanium experiment we adopt a fiducial detector mass of 1 ton and an exposure of three years. Taking into account the percentage of events that survive the selection cuts $\eta_\text{cut}$ and the nuclear recoil acceptance for germanium $A_{NR} = 90\%$ the effective exposure is $\epsilon_\text{eff} = 2.16$ ton$\times$years.

Several sources of background can induce additional recoil events in direct detection experiments, such as cosmic rays, or radioactive contaminations. Future detectors will apply a variety of advanced techniques in order to achieve extreme radio-purity and self-shielding of the detector, minimisation of cosmic ray events and precise determination of charge-to-light and charge-to-phonon ratios, in order to limit the background to $<1$ event per effective exposure. Given these prospects in the following we assume that backgrounds are negligible.

We do not include the energy resolution of the detectors, as for both target materials including energy resolution smearing has a negligible impact on the recoil rate, except possibly near threshold.  The scenario considered here is therefore somewhat idealised, which means that the statistical uncertainties we identify are unavoidable, inherent to the WIMP benchmark point and target exposure, rather than a reflection of systematic uncertainties in detector response, backgrounds or modelling of the dark matter halo.


\section{Statistical methodology}\label{statsection}

\subsection{Mock data generation}

The data set for a direct dark matter experiment consists of the total number of observed events $\hat{N}_R$ and the spectrum of recoil energies $\{\hat{E}_R^i\}$, with $i = 1,..,\hat{N}_R$. The likelihood function $\like (\params)$ for the WIMP parameters $\params = \{ m_\chi, \sigma_p^{SI} \}$ is given by the Poisson probability of observing $\hat{N}_R$ events, multiplied by the probabilities of each event of energy ${E}_R^i$ having been drawn from the predicted probability distribution of event energies $P(E_R | \theta)$
\begin{equation}
\label{unbinned}
\like(\params)=\frac{N_R(\params)^{\hat{N}_R}}{\hat{N}_R!}\exp\left[-N_R(\params)\right] \prod_{i = 1}^{\hat{N}_R} P(\hat{E}_R^i |\theta) \quad.
\end{equation}
Notice that in the above we have replaced the (latent, unobserved) true recoil energy ${E}_R^i$ by the observed value $\hat{E}_R^i$, thus assuming that energy resolution of the detectors is negligible, as outlined in the previous section.  $N_R(\params)$ can be computed from Eq.~\eqref{nocounts}, using the experimental characteristics in Table \ref{experiments}.  The distribution $P(\hat{E}_R,\theta)$ is no more than the normalized recoil spectrum
\begin{equation}
\label{unbinned2}
P(\hat{E}_R,\theta) = \frac{dR / dE_R (\hat{E}_R,\params)}{\int_{E_\text{min}}^{E_\text{max}} dE^{\prime}_R dR / dE^{\prime}_R (E^{\prime}_R,\params) } \quad,
\end{equation}
where the rate $dR / dE_R (E_R,\params)$ is given in Eq.~\eqref{recoil2}. Note that the efficiency parameter $\epsilon_{\mathrm{eff}}$ drops out in the one-event likelihood because we assume that this function is independent of recoil energy. For both the Xe and the Ge target the integration limits are $E_\text{min} = 10$ keV and $E_\text{max} = 100$ keV. As explained in the previous section no background events are included in $\hat{N}_R$, as we assume the background to be negligible. The so-called unbinned likelihood function in Eq.~\eqref{unbinned} has been employed by both the XENON and the CDMS collaborations~\cite{Aprile:2011hx,Ahmed:2008eu}. The likelihood function for the combined data set of our two toy experiments is given by the product of the individual likelihood functions, each found from Eq.~\eqref{unbinned}.

The mock data sets for the experiments are generated as follows. First, the measured total number of counts $\hat{N}_R$ is drawn from a Poisson distribution with mean equal to the benchmark number of counts $N_R$. Then, values for the measured recoil energies $\{\hat{E}_R^i\}$, $i = 1, .., \hat{N}_R$ are drawn from the differential event rate $dR/dE_R (E_R)$, given in Eq.~\eqref{recoil2}, for the benchmark value of the parameters. 

\subsection{Parameter reconstruction technique}

We employ Bayesian methods to scan over the parameter space and reconstruct the WIMP properties, see~\cite{Trotta:2008qt} for further details. The cornerstone of Bayesian parameter inference is Bayes' theorem
\begin{equation} \label{eq:bayes}
p(\params | \data) = \frac{\mathcal{L}(\params) p(\params)}{p(\data)} \quad,
\end{equation}
where $p(\params | \data)$ is the posterior probability density function (pdf), $\mathcal{L}(\params)$ is the likelihood function and $p(\params$) is the prior distribution on the parameters. The evidence is given by $p(\data)$, which in the context of parameter inference acts as a normalisation constant and will not be of interest in the following. There are two possible ways of looking at parameter inference: either in the Bayesian context (where the posterior pdf is the relevant quantity) or in the frequentist framework (where the likelihood function or a related test statistic is considered). In this work, we will use Bayesian Markov Chain Monte Carlo (MCMC) techniques to obtain samples from the posterior pdf of Eq.~\eqref{eq:bayes}, but we will also use these samples to map the likelihood function in the parameter space of interest, here the WIMP mass and the WIMP-proton spin-independent scattering cross-section, $\params = \{m_{\chi},\sigma_p^{SI}\}$. In order to sample from the posterior distribution on these parameters, we have to specify their prior pdf $p(\theta)$. Without assuming a specific underlying WIMP model there are no a priori constraints on $m_{\chi}$ and $\sigma_p^{SI}$. Therefore, we choose uniform priors on the log of both the WIMP mass and cross-section, reflecting ignorance on their order of magnitude. The mass prior range is fixed to $1\le\log_{10}(m_{\chi}/\textnormal{GeV}) \le 3$. The range of the cross-section prior is chosen to span two orders of magnitude around the benchmark cross-section.  We extend this range where required, to avoid regions of high posterior probability density touching the prior boundary.
 
Because the likelihood function is unimodal and well-behaved, and the parameter space is of low dimensionality ($D = 2$), we can efficiently sample the posterior pdf using MCMC methods and use the ensuing samples to map out the likelihood function in a quasi-frequentist sense (see~\cite{Feroz:2011bj} for a detailed study of profile likelihood evaluation using Bayesian techniques in the context of supersymmetric models). To this end, we use a Metropolis-Hastings algorithm~\cite{metropolis:1087,HASTINGS01041970} to generate a ``chain'' of samples from the posterior pdf.  As our proposal distribution we take a two-dimensional Gaussian centred on the previous point in the chain; its covariance matrix is chosen according to earlier test runs. For some of the benchmark points we consider, the shape of the posterior distribution can vary strongly because of statistical fluctuations in the data realisation. In these cases, to achieve an efficient and complete sampling of the posterior we adopt a mixture strategy MCMC: our proposal distribution is a mixture of two different two-dimensional Gaussians, whose covariance matrices are chosen (from earlier test runs) to match the two very different shapes of the posterior distribution that can arise from the same benchmark model due to statistical fluctuations in the data (``good'' reconstructions and ``bad'' reconstructions, to be defined more precisely below). Every third proposal of the MCMC is not drawn from this Gaussian mixture, but instead is taken in a random direction, with a step size tuned to achieve an acceptable efficiency, in order to protect against under-exploration of the tails of the posterior.
 
Each Markov chain contains a minimum number $N = 3 \times 10^5$ samples; this ensures high enough statistics for a successful coverage investigation. Some benchmark models lead to a very spread-out posterior distribution. In these cases we further increased the number of points in the chains, up to a maximum of $N = 5 \times 10^5$ points. We discarded the initial $10^4$ samples of each chain (the so-called ``burn-in''). We checked that this is sufficient to ensure that the resulting distribution is independent of the starting point of the MCMC and that the results of our analysis are stable when the length of the chains is doubled. Finally, we tested our MCMC method on toy models with known analytic posterior distributions, in order to verify its suitability and numerical stability. 

\subsection{Coverage}

There are two ways of reporting inferences: $x\%$ credible intervals (Bayesian) contain a fraction $x$ of the posterior probability; they express the posterior degree of belief about the value of the parameter considered after the data and any prior information have been taken into account. An $x\%$ confidence interval (Frequentist) is built from the likelihood function alone, and, ideally, it ought to contain (``cover'') the true value of the parameter  x\% of the time, when repeatedly applied to mock data generated from those true parameter values. This requirement leads to the concept of ``coverage''. Coverage is an inherently frequentist concept, and it is not necessarily of concern to Bayesian statistics, although reliable behaviour of Bayesian credible intervals under repeated sampling is arguably also a desirable property. In the following, we will mainly focus on evaluating the coverage and other statistical properties of (frequentist) confidence intervals, for the reasons outlined below. 

The profile likelihood test statistic for a point $X$ in some $N$-dimensional subspace $\Theta_N$ of the full $M$-dimensional parameter space $\Theta_M$ (i.e. $X\in\Theta_N\subset\Theta_M$), is
\begin{equation} \label{eq:teststatistics}
\lambda(X) = -2 \ln \left(\frac{\like[X, \hat{\Theta}_{M-N}(X)]}{\like_\text{max}}\right) \quad.
\end{equation}
Here $\like_\text{max}$ is the unconditional maximum likelihood i.e. the global maximum likelihood value across the entire $M$-dimensional parameter space.  $\like[X, \hat{\Theta}_{M-N}(X)]$ is the conditional maximum likelihood for the given point $X$.  The subspace $\Theta_{M-N}$ refers to the section of $\Theta_M$ that is not spanned by $\Theta_{N}$.  $\hat{\Theta}_{M-N}(X)$ is the conditional maximum likelihood estimate of the values of the parameters in $\Theta_{M-N}$ for $X$, i.e. the specific combination of the other $M-N$ parameters that maximises the likelihood for the chosen $X$ in $\Theta_N$.  Confidence intervals with exact coverage can always be constructed by Monte Carlo evaluation of the distribution of $\lambda(X)$, as described in Ref.~\cite{PhysRevD573873}, 
but in practice this may be a complicated and time-consuming procedure.

Wilks' theorem \cite{Wilks1938} shows that under certain regularity conditions, Eq.~\eqref{eq:teststatistics} converges asymptotically to a chi-square distribution with $N$ degrees of freedom.
Assuming Wilks' theorem holds, it is simple to define confidence intervals using the profile likelihood function and standard lookup tables for the chi-square distribution. However, in practice there is no guarantee that such confidence intervals will have the desired coverage properties, especially in cases where the likelihood function is strongly non-Gaussian, which leads to a lack of convergence of the test statistic to its asymptotic behaviour. Under-coverage (over-coverage) of a confidence interval means that the interval is too short (too large). While over-coverage is unnecessarily conservative, under-coverage can be a particularly severe problem, as the true value of the parameters will lie outside the stated interval a larger fraction of the time than its stated confidence level implies. 

In the following analysis we discuss the coverage of Wilks-based 1D confidence intervals for the WIMP mass and spin-independent cross-section. The profile likelihood is constructed by binning the 2D parameter space ($\{m_{\chi},\sigma_p^{SI}\}$), and determining the test statistics~\eqref{eq:teststatistics} in each bin. We then use Wilks' theorem to find the confidence level of interest. We used 750 bins in each direction of parameter space, choosing the bin size so that they covered the whole range spanned by the samples.  We found that a significantly larger number of bins leads to large numerical noise, while a smaller number gives too coarse a likelihood mapping and hence artificial over-coverage (as tested on Gaussian toy models, for which the coverage is exact).

\subsection{Performance of parameter reconstruction} \label{statstests}

In addition to determining how well the Wilks-based confidence levels cover the benchmark models, we are interested in estimating how well one may expect to constrain WIMP properties from future direct detection data sets, {\em including realisation noise}.  An important indicator is the uncertainty in the reconstructed parameters. In order to quantify this, we consider the expected fractional uncertainty (e.f.u.) along a direction in parameter space. The fractional uncertainty (f.u.) is defined as the fractional length of the $68 \%$ confidence interval relative to the benchmark parameter value $\theta_\mathrm{true}$:
\begin{equation}
\textnormal{f.u.} = \frac{\theta^{68\%}_{\textnormal{max}} - \theta^{68\%}_{\textnormal{min}}}{\theta_\mathrm{true}} \quad.
\end{equation}
The e.f.u.\ is the average of this quantity over 100 reconstructions. However, even a benchmark model with a small {\em average} f.u. may contain a sizeable number of reconstructions with a large parameter uncertainty. Therefore, in addition to the e.f.u.\ we also count the number of `bad' reconstructions in 100 reconstructions. A bad case is defined as a reconstruction with an f.u.$> 0.75$, in which case only very limited constraints can be placed on the parameter in question ($m_\chi$ or $\sigma_\mathrm{SI}$) from the data.

The f.u. is somewhat similar to the statistical quantity known as effect size \cite{Cohen88,Bausell06}, which for the case of $\sigma_\mathrm{SI}$ is  
\begin{equation}
d \equiv \frac{(\hat{\sigma}_\mathrm{SI} - \sigma_{\mathrm{SI,null}})}{SD} \quad.
\end{equation}
Here $\hat{\sigma}_\mathrm{SI}$ and $SD$ are the mean and standard deviation, respectively, of a series of repeated measurements of $\sigma_\mathrm{SI}$.  In our case, an equivalent role to $\hat{\sigma}_\mathrm{SI}$ and $SD$ are played by the best-fit reconstructed value of $\sigma_\mathrm{SI}$, and half the width of the corresponding 68\% CI.  
This is because these quantities are good estimators for, respectively, the true value of $\sigma_\mathrm{SI}$ and the standard deviation of $\hat{\sigma}_\mathrm{SI}$, the observed best-fit value.
The quantity $\sigma_{\mathrm{SI,null}}$ refers to the value of $\sigma_\mathrm{SI}$ under the null hypothesis, i.e.\ the default situation against which the effect is being sought.  In our case, the null hypothesis is simply that there is no WIMP signal, so $\sigma_\mathrm{SI}=0$. Therefore, in the limit of zero bias, where the best-fit value of $\sigma_\mathrm{SI}$ is exactly equal to the benchmark value, e.f.u.\ is approximately equivalent to $2d^{-1}$.  The case of WIMP mass is less straightforward, as $m_\chi$ is undefined under the null hypothesis.

One of the basic properties of statistical inference is that the power of a statistical test (its ability to avoid excluding a true hypothesis that differs from the null hypothesis) increases with $d$ \cite{Murphy04,Bausell06}.  This is simply the statement that larger effects can be detected more easily.  We can therefore see that the e.f.u.\ not only relates to the precision with which the WIMP mass can be reconstructed, but also gives some idea of the statistical power for detection of a WIMP with this mass.  That is, a smaller e.f.u.\ indicates that a model can be detected more easily, so we expect the e.f.u.\ to roughly track the sensitivity of an experiment across the WIMP parameter space.

We can further investigate the performance of the statistical reconstruction by explicitly considering the bias\footnote{Another useful quantity is the so-called ``mean squared error''
(MSE) for the parameters, given by the sum of the bias squared and the variance. We have found that the MSE behaves qualitatively similarly to the e.f.u., so we do not discuss it separately.} for the parameters $m_{\chi}$ and $\sigma_p^{SI}$. The statistical bias for a parameter $\params$ is the expectation value of the difference between the best fit value $\hat{\params}_\text{bf}$ resulting from the reconstruction and the true value $\params_\text{true}$, i.e.
\beq
\textnormal{bias} = \left\langle \hat{\params}_\text{bf} - \params_\text{true} \right\rangle \quad .
\eeq
As for the e.f.u., the expectation is taken by averaging the observed bias over 100 reconstructions.  In the following we focus on the e.f.u.\ and bias of the reconstructed WIMP mass, as the performance of the reconstruction is expected to typically be poorer in the mass than the cross-section direction, due to the impact of statistical fluctuations on the observed recoil spectrum.

\section{Results}\label{results}

\subsection{The impact of statistical fluctuations on the reconstruction}

\begin{figure*} 
  \centering
\includegraphics[width=0.495\linewidth,trim = 30mm 80mm 30mm 80mm]{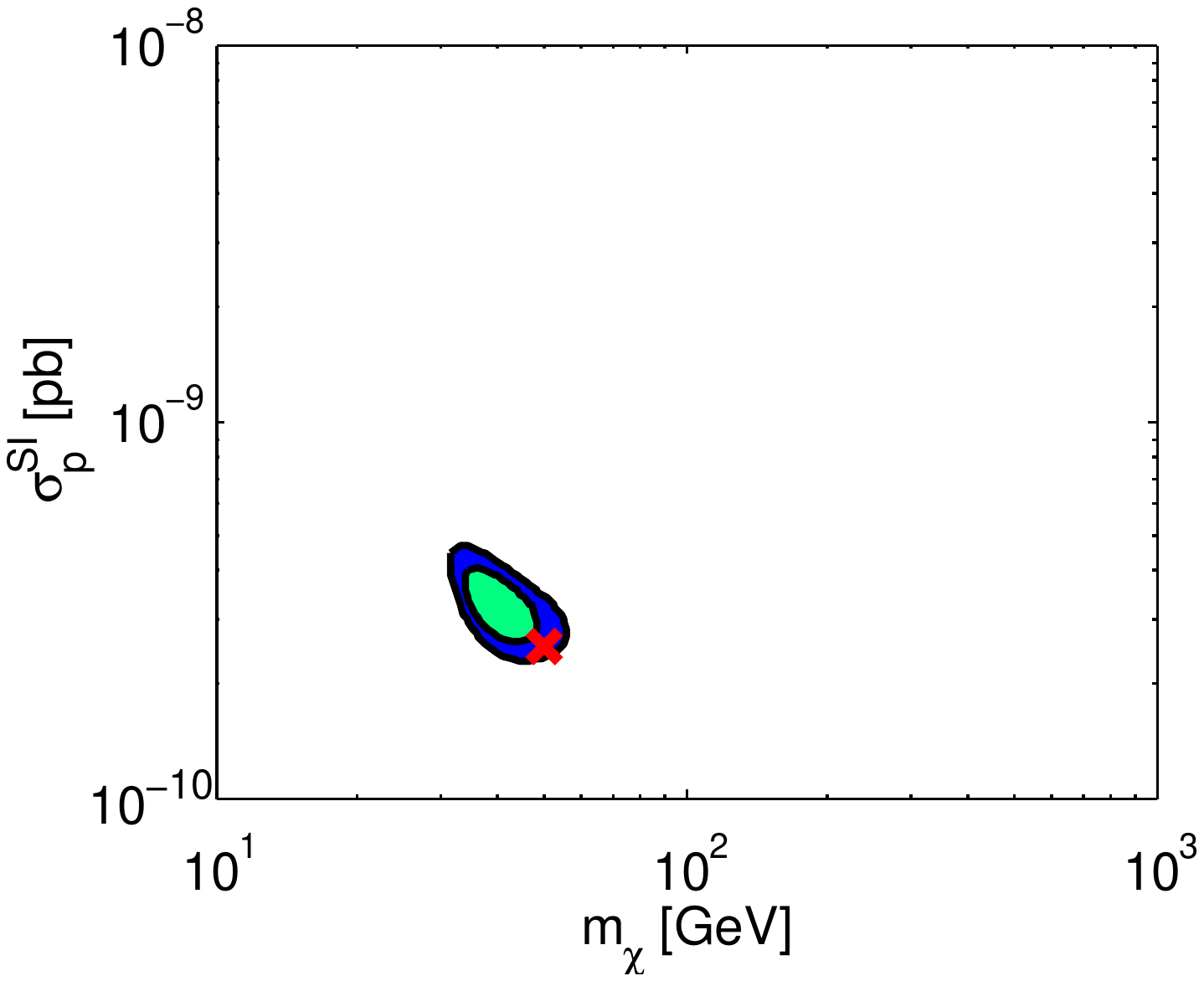} 
\includegraphics[width=0.495\linewidth,trim = 30mm 80mm 30mm 80mm]{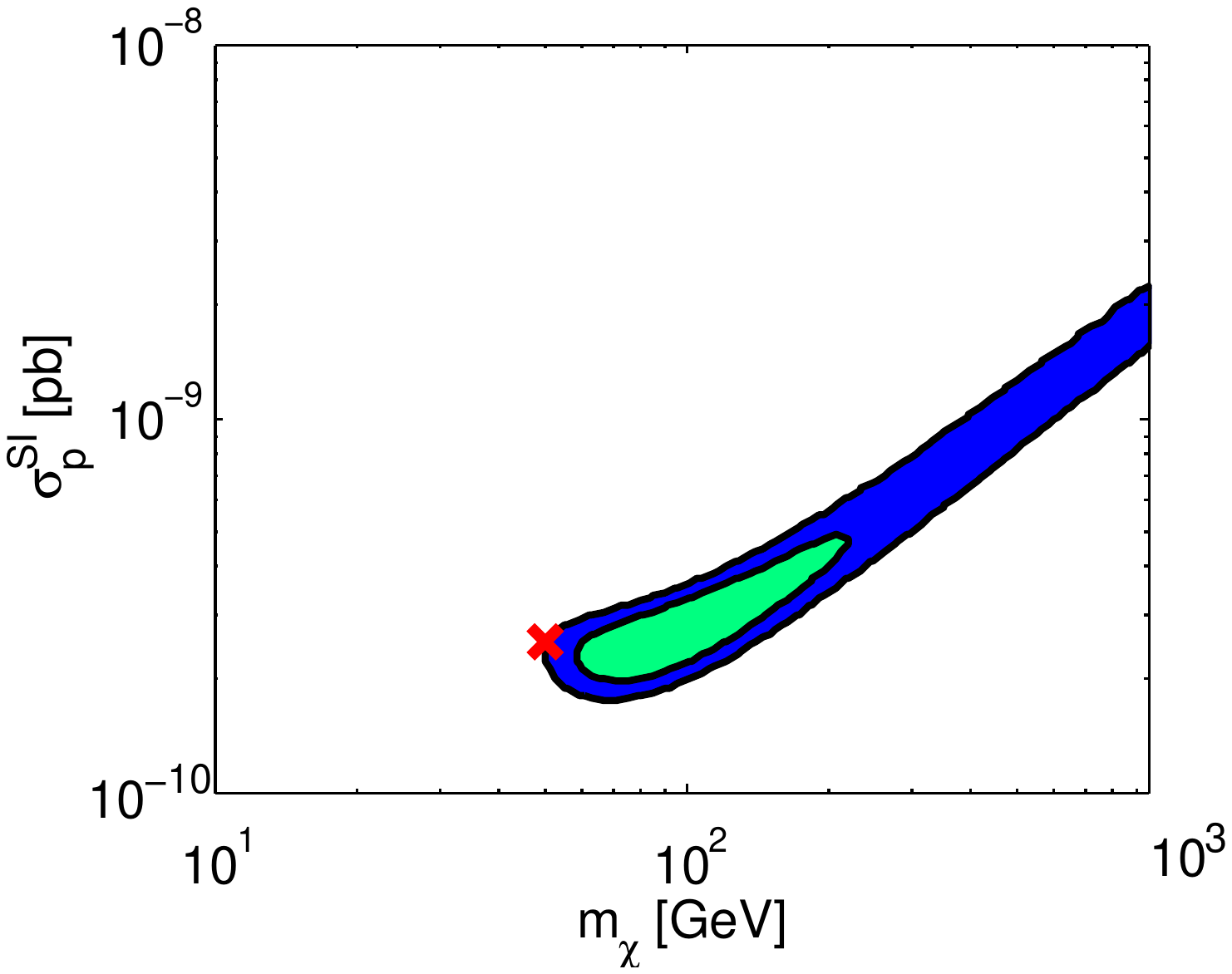} \\
\includegraphics[width=0.485\linewidth,trim = 10mm 70mm 10mm 70mm]{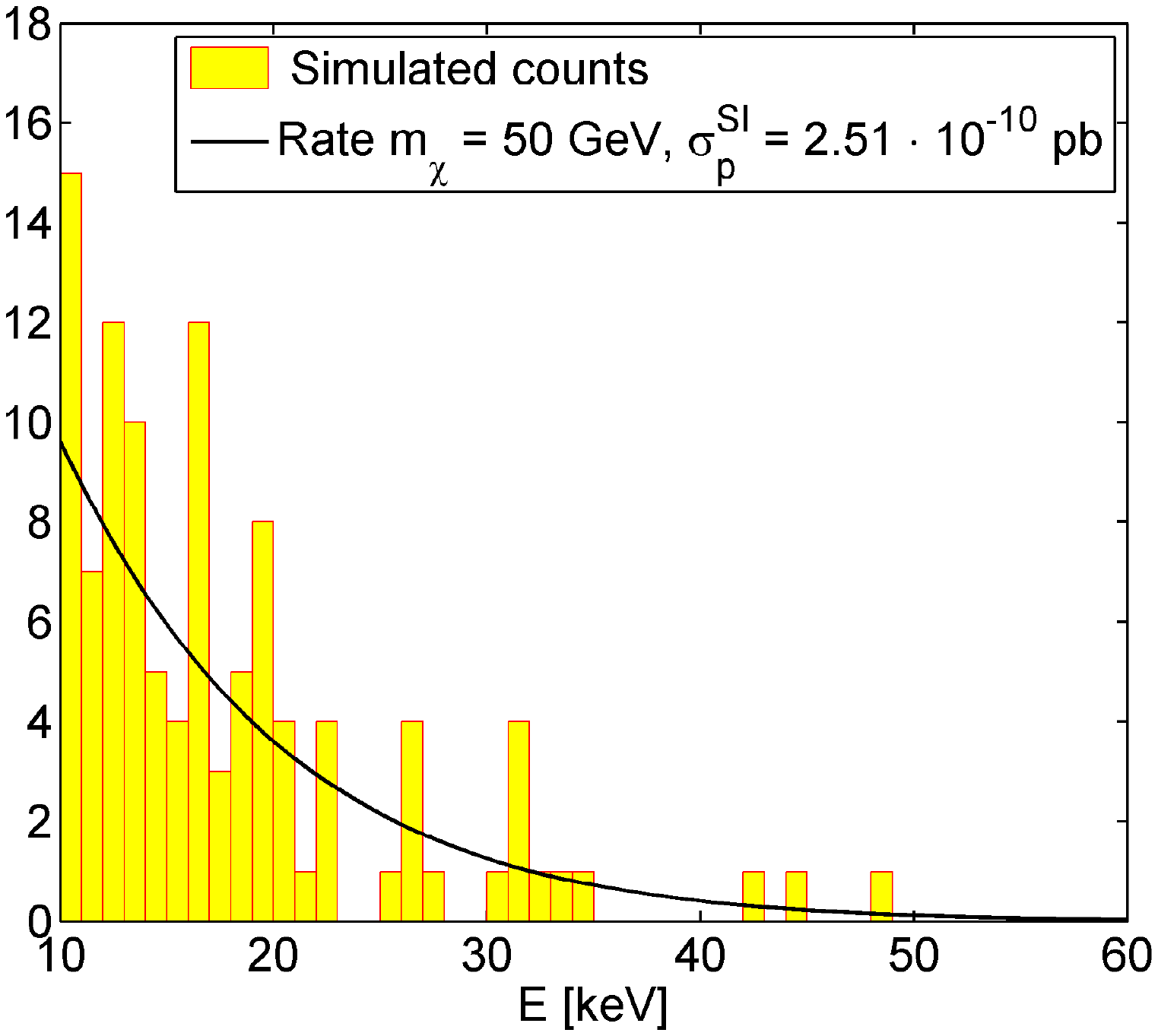}
\includegraphics[width=0.485\linewidth,trim = 10mm 70mm 10mm 70mm]{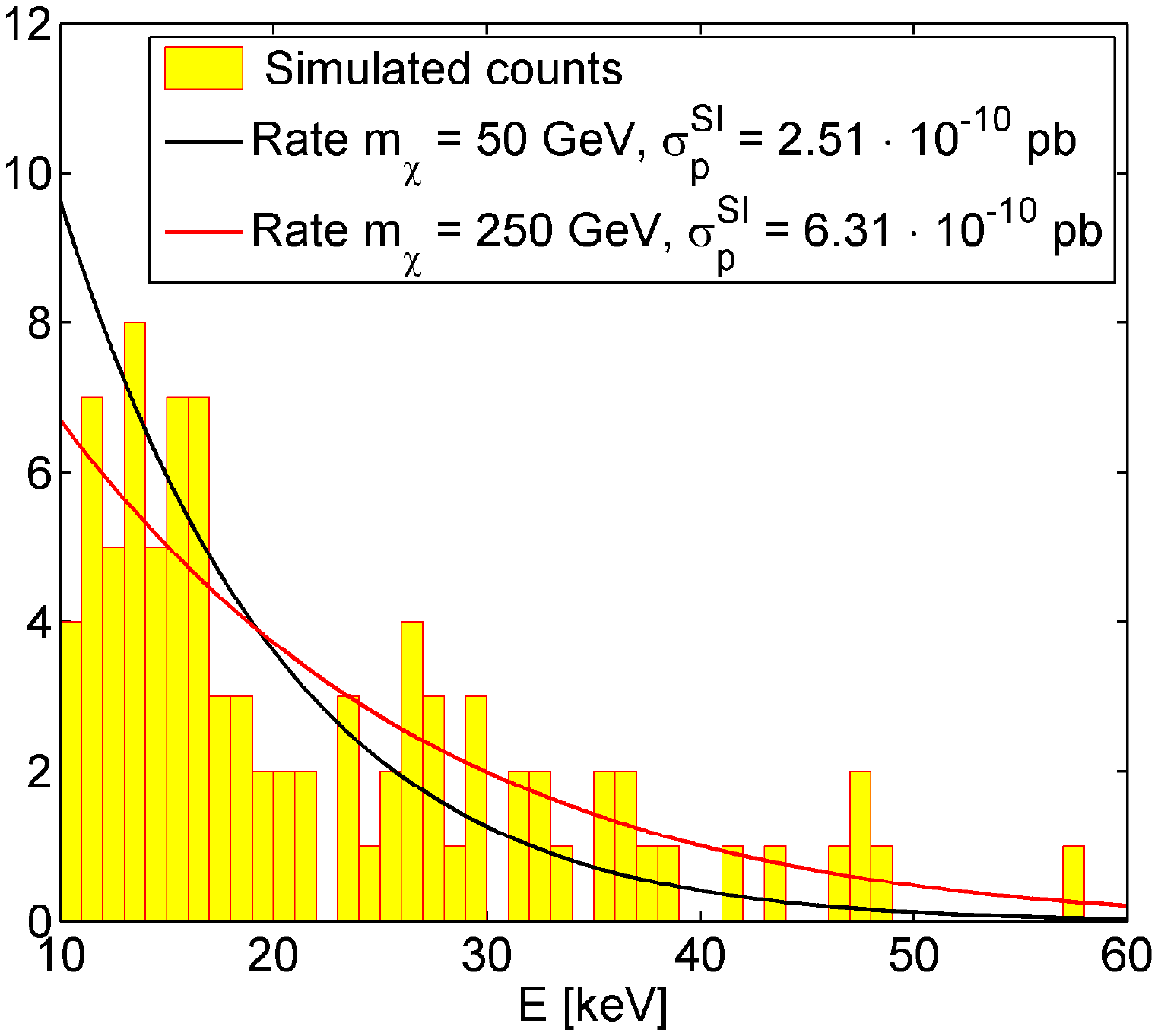} 
\caption{The left (right) panels show examples for a good (bad) reconstruction of the WIMP benchmark model with true values $m_{\chi} = 50$ GeV, $\sigma_p^{SI} = 2.51 \times 10^{-10}$. The difference is exclusively in statistical fluctuations in the simulated data. Top panels: $68.3\%$ and $95.4\%$ confidence levels in the $m_{\chi} - \sigma_p^{SI}$ plane; the red cross shows the true value. Bottom panel: energy spectrum of the mock data (yellow histogram - recall that we use an unbinned likelihood function, the counts are binned for a better visualization), true rate $dR/dE (E)$ (black) and for the ``bad'' reconstruction an example of a rate (red) with a higher likelihood than the true rate.}
  \label{bm2fig} 
\end{figure*}

We investigate the performance of the reconstruction of WIMP properties for six benchmark masses $m_{\chi} = \{25, 35, 50, 70, 100, 250\}$ GeV, and six spin-independent WIMP-proton cross-sections $\sigma_p^{SI} = \{1.00 \times 10^{-8},3.98 \times 10^{-9}, 1.58 \times 10^{-9}, 6.31 \times 10^{-10}, 2.51 \times 10^{-10}, 1.00 \times 10^{-10}\}$ pb, thus 36 benchmark models in total. The number of dark matter recoil events above threshold for our Xe experiment (see section \ref{exp}) for these benchmark points is in the range $10 \lsim N_R \lsim 4000$. As we focus on the case of a significant detection in a future experiment, we do not investigate the statistical properties of benchmark points in the very low counts regime, where $N_R < 10$, as it is hard to constrain much of anything with fewer than $\sim 10$ events.

Before we present results for our coverage study and the quantitative description of the performance of parameter estimation, we show examples of good and poor reconstructions of WIMP parameters based on the mock data sets of a specific benchmark point.  These examples illustrate points that will be important in our coverage and performance studies.

Two examples of the reconstruction using Xe data are shown in Fig.~\ref{bm2fig} for a benchmark model with WIMP mass $m_{\chi} = 50$ GeV and spin-independent WIMP-proton cross-section $\sigma_p^{SI} = 2.51 \times 10^{-10}$ pb. This is an example of a benchmark point for which the performance of the reconstruction can vary strongly with the mock data.  We show on the left of Fig.~\ref{bm2fig} an example of a ``good'' reconstruction (i.e., well constrained likelihood in the $m_{\chi} - \sigma_p^{SI}$ plane), and on the right of Fig.~\ref{bm2fig} an example of a ``bad'' reconstruction (leading to an essentially unconstrained likelihood).  For both cases we show the $68.3\%$ and $95.4\%$ likelihood contours (top) and the energy spectrum of the mock events (bottom), compared with the theoretical spectrum of the benchmark model (shown in black). 

For the first example (left) both the $68.3\%$ and the $95.4\%$ confidence level spans a small range of masses and the benchmark point is well reconstructed. The distribution of the observed energies agrees well with the true benchmark rate. In contrast, the second example (right) leads to confidence levels that spread over a large mass range; at $95.4\%$ confidence only a lower limit on the WIMP mass can be inferred (note that the $95.4\%$ contour does not close, but is cut off at the upper mass prior limit $m_{\chi} = 1000$ GeV).  The benchmark point is badly reconstructed mostly because of the presence of a relatively large number of high-energy counts at $E > 40$ keV. Events with these energies are an unlikely realisation of the benchmark WIMP spectrum, but can appear in the data due to statistical fluctuations. Poisson noise has flattened the observed energy spectrum relative to the predicted energy spectrum. The confidence intervals show ``runaway'' behaviour towards high mass because a flat energy spectrum is indicative of high masses, and the energy spectra for $m_\chi \gg m_N$ are nearly identical.  As an example, the theoretical spectrum for a WIMP model with $m_{\chi} = 250$ GeV, $\sigma_p^{SI} = 6.31 \times 10^{-10}$ pb is shown in red in the bottom right panel. Clearly this model is a better fit to the simulated events than the benchmark model. 

Note that this benchmark model leads to a large number of events ($N_R \sim 100$), so that one would naively expect that statistical fluctuations in the realised spectrum ought to have a minor impact. This is clearly not the case, as the bad reconstruction in the right panels of Fig.~\ref{bm2fig} shows that even with $\sim$100 events, the parameter reconstruction can be poor.  Even though we show in the rest of this section that this benchmark is relatively well-behaved---the coverage is exact for most intervals, the e.f.u. and bias are low, and the expected number of large-f.u. outliers is fairly small---there is a non-negligible probability that particular realisations of data sets for this benchmark lead to catastrophically poor WIMP parameter reconstructions.

\subsection{Results from the coverage analysis}

\begin{figure*}
  \centering
\includegraphics[width=0.495\linewidth,trim = 10mm 70mm 10mm 70mm]{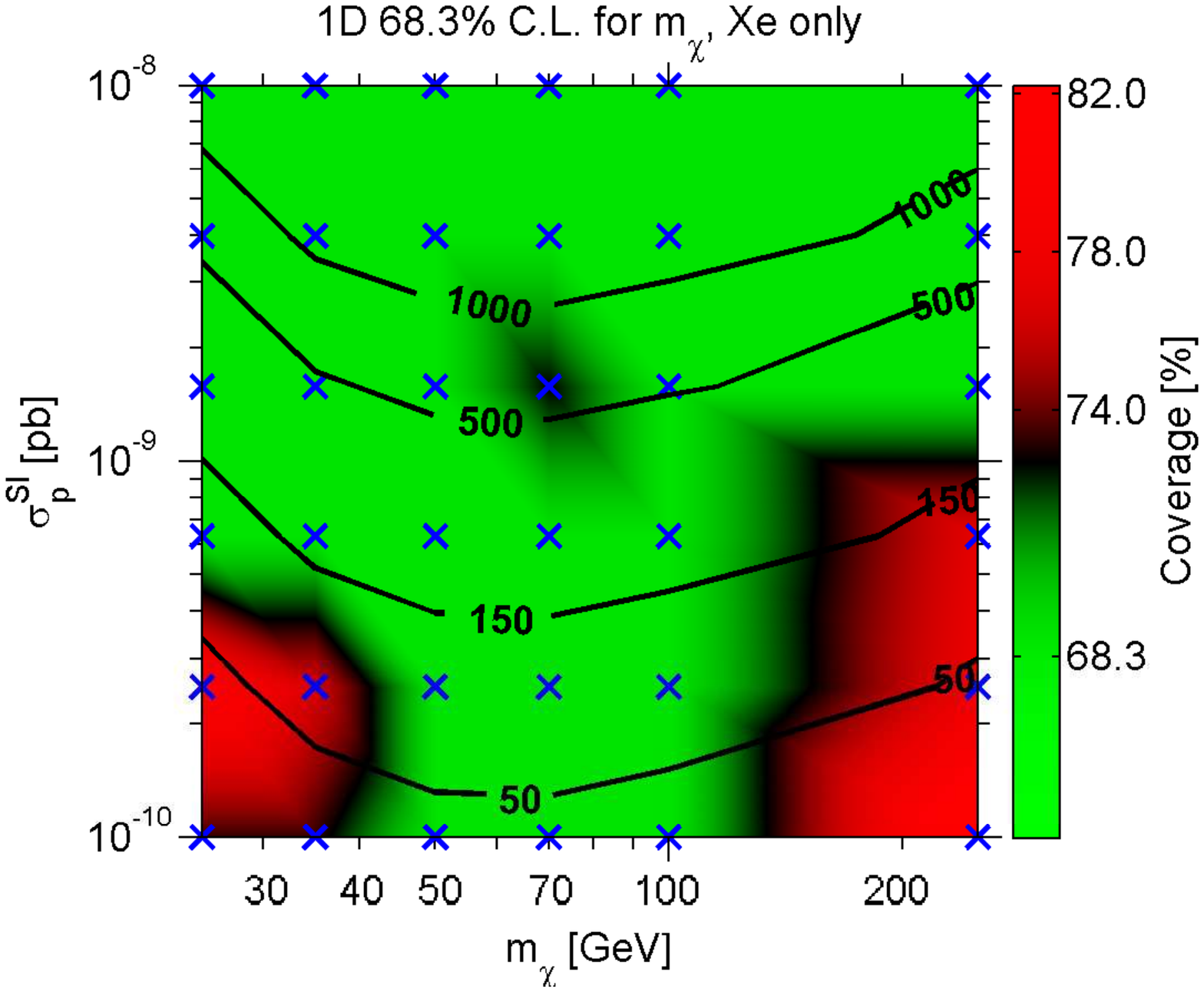} 
\includegraphics[width=0.495\linewidth,trim = 10mm 70mm 10mm 70mm]{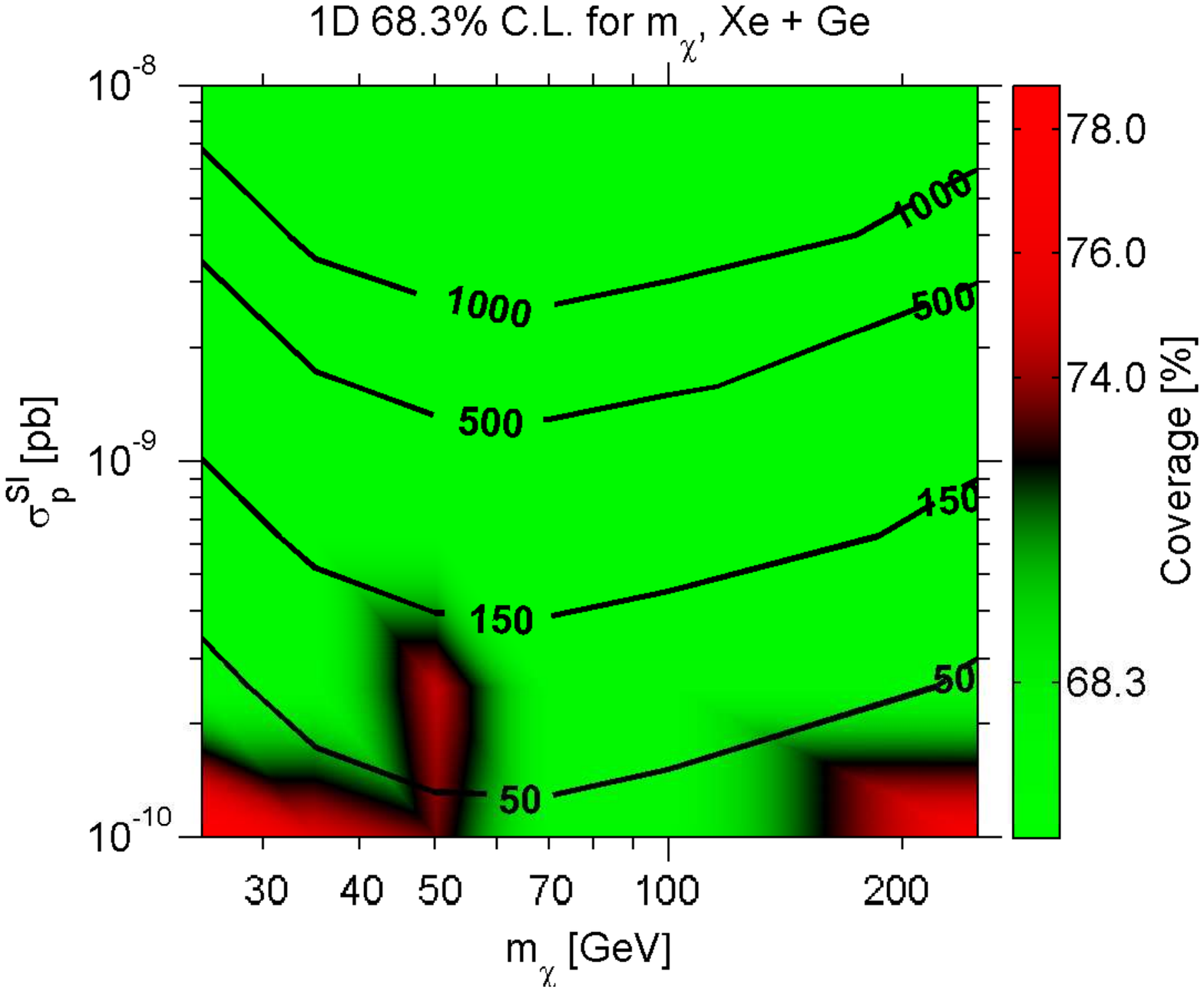} \\
\includegraphics[width=0.495\linewidth,trim = 10mm 70mm 10mm 70mm]{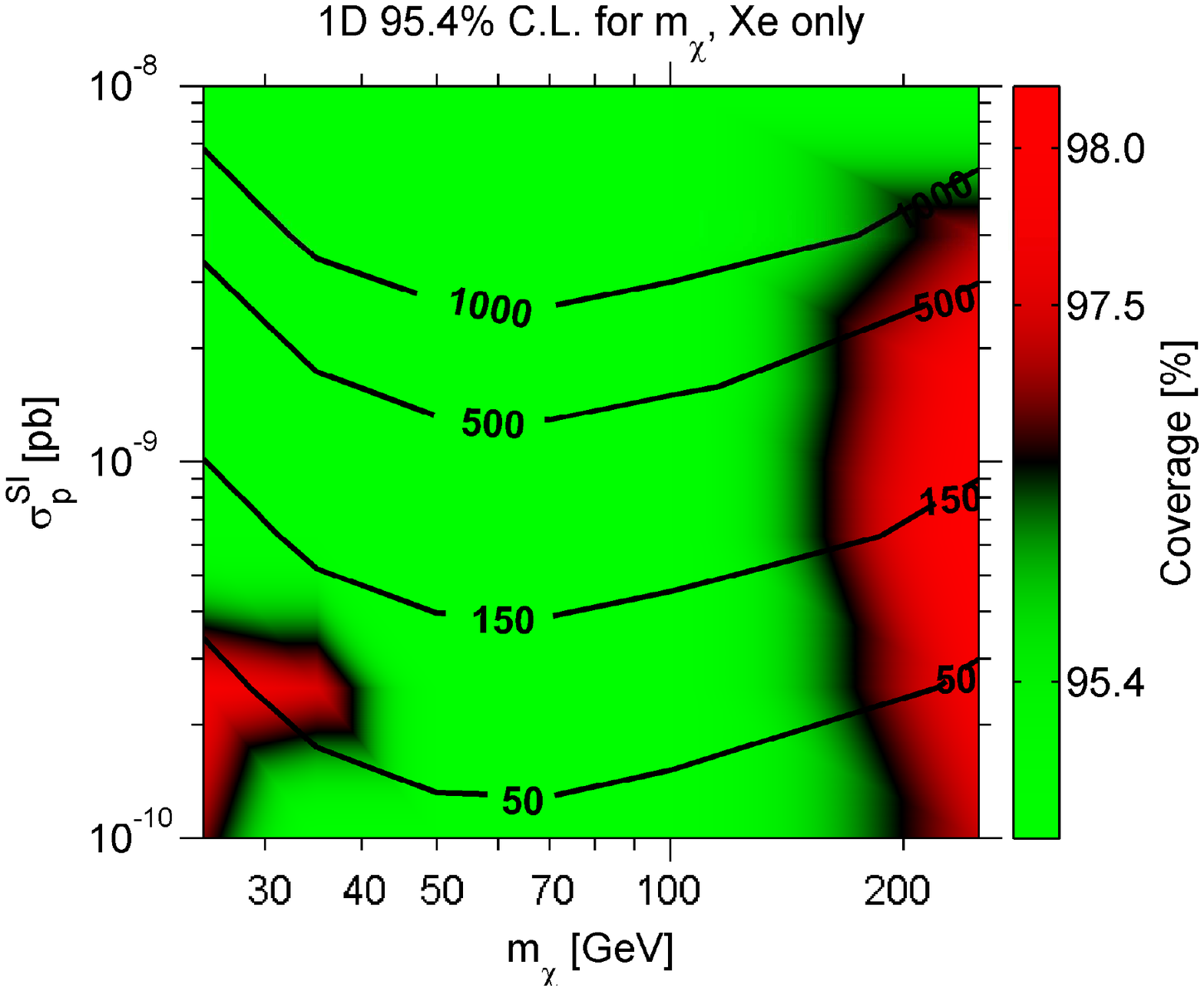} 
\includegraphics[width=0.495\linewidth,trim = 10mm 70mm 10mm 70mm]{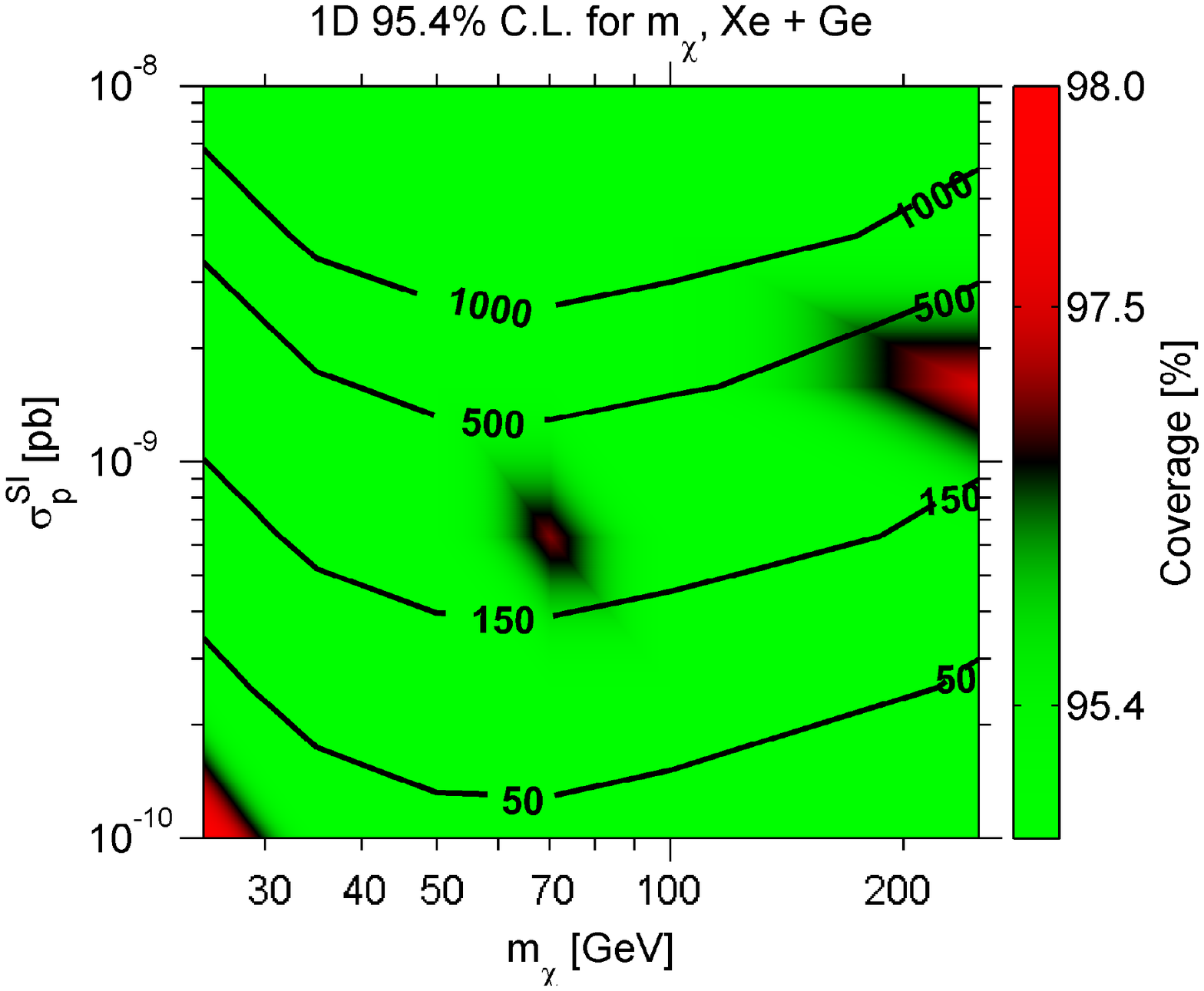} 
\caption{Coverage results for the 1D $68.3\%$ (top) and $95.4\%$ (bottom) confidence interval for the WIMP mass in the $m_{\chi} - \sigma_p^{SI}$ plane, for simulated Xe target (left) and for a combination of Xe+Ge (right). Green (red) regions show ``exact'' coverage (over-coverage), as defined in the text. Black regions correspond to a transition from exact coverage to over-coverage. No under-coverage is observed. Isocontours of the expected number of counts in the Xe experiment are given in black. In the upper left plot, the benchmark points studied are indicated by blue crosses. The `flares' pattern seen in some points are an artefact of the interpolation scheme used to generate the plots.} \label{1Dm}
\end{figure*}

In order to investigate the coverage results for the 1D $68.3\%$ and $95.4\%$ confidence intervals for $m_{\chi}$ and $\sigma_p^{SI}$, for both Xe data and a combination of Xe+Ge date, we generate 1000 mock data sets for each of the 36 benchmark models, as outlined in section \ref{statsection}. The 1D $68.3\%$ ($1 \sigma$) and $95.4\%$ ($2 \sigma$) confidence levels are constructed using Wilks' theorem and we count how often the true value of the WIMP mass and cross-section are found within the stated CL. We further subdivide the 1000 reconstructions into 10 subsets, of 100 reconstructions each, and we compute the coverage for each subset. We take the standard error of these ten values to estimate the statistical error of our coverage analysis, encompassing the uncertainty coming from finite numerical samples of the likelihood and the finite number of reconstructions. Although this statistical error on the coverage value varies mildly across benchmark points, it is sufficient for our purposes to use its average over all benchmark points. This leads to an estimated $1\sigma$ error of $4.5\%$ for the $68.3\%$ intervals, and of $1.9\%$ for the $95.4\%$ intervals. 

We start by discussing the 1D $68.3\%$ and $95.4\%$ confidence intervals for $m_{\chi}$, shown in the top and bottom panels of Fig.~\ref{1Dm}, respectively. On the left-hand side we show the coverage results obtained for a Xe target, on the right-hand side we show results for the combined data set Xe+Ge. From the above estimate of the error on the coverage, we define the coverage to be ``exact'' if it lies in the range $(63.8,72.8)\%$ and $(93.5,97.3)\%$ for the $68.3\%$ and $95.4\%$ contours, respectively. Benchmark points showing ``exact'' coverage within errors are displayed in green. Coverage values $>72.8\%$ ($>97.3\%$) correspond to over-coverage and are shown in red. Coverage values $<63.8\%$ ($<93.5\%$) correspond to under-coverage. However, none of the benchmark points studied here leads to under-coverage of any of the confidence intervals. Benchmark points at the upper boundary of exact coverage or the lower boundary of over-coverage are displayed in black. For reference, isocontours of the expected number of counts $N_R$ in a Xe experiment are also shown. 

\begin{figure*}
  \centering
\includegraphics[width=0.65\linewidth,trim = 10mm 65mm 10mm 65mm]{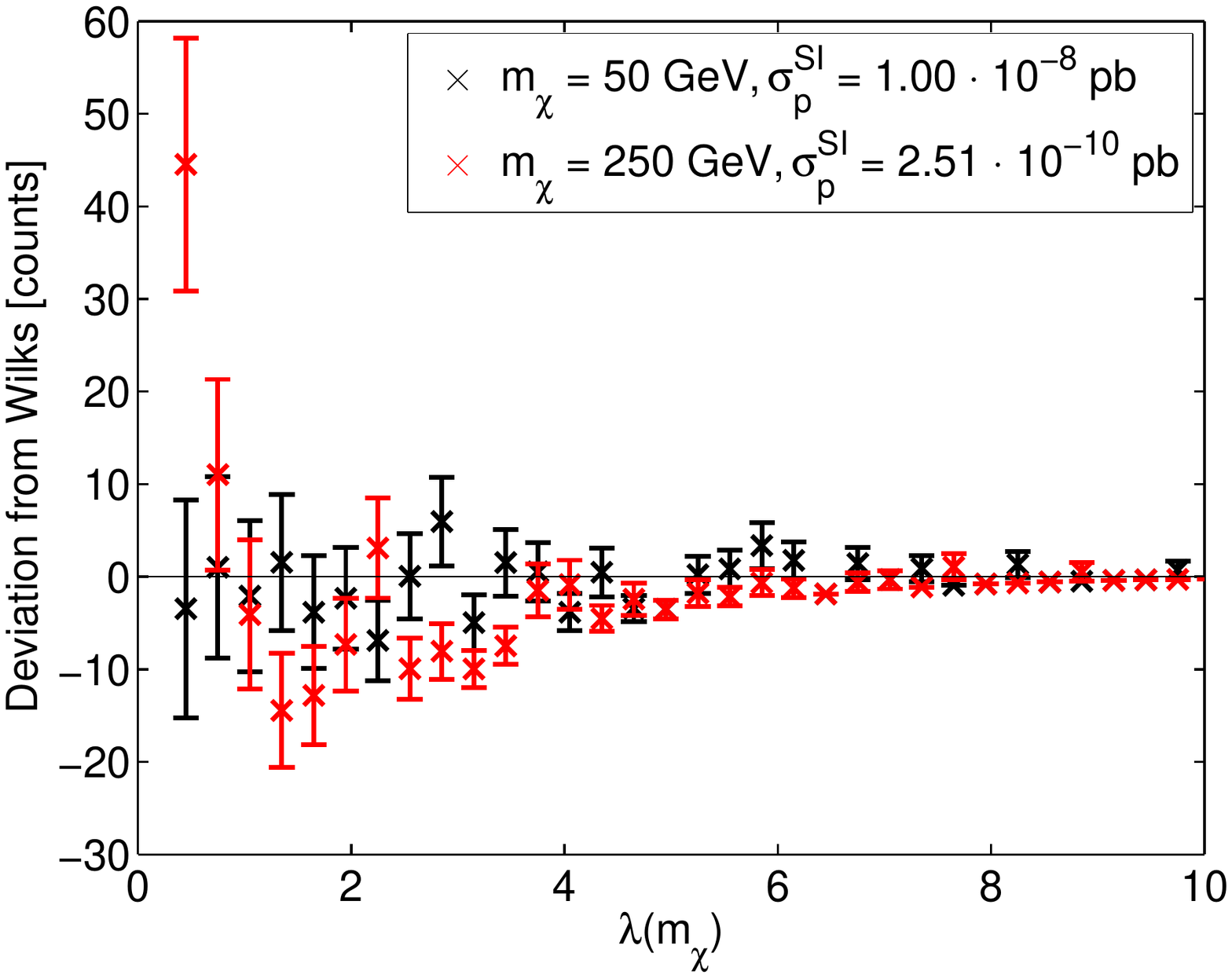} 
\caption{Difference between the histogram of the profile likelihood test statistic $\lambda(m_{\chi})$ from mock data sets and the value of the chi-square distribution with 1 degree of freedom (as predicted by Wilks' theorem) at the centre of each bin, as a function of $\lambda(m_{\chi})$, for two different WIMP benchmark points.. This difference quantifies the deviation from Wilks' theorem for these two benchmark points. For each benchmark point, 10$^3$ realisations of mock data sets have been used to construct this histogram. Errorbars assume Poisson count statistics.}\label{Fig:Wilks}
\end{figure*}

For the Xe-only case, we find that most benchmark points lead to exact coverage of the 1D $68.3\%$ and $95.4\%$ contours. For the $68.3\%$ interval there is a region observed at high cross-sections and intermediate WIMP masses that borders on over-coverage; this is most likely the result of a statistical fluctuation. For both the $68.3\%$ interval and the $95.4\%$ interval, two regions leading to significant over-coverage can be identified, one at large $m_{\chi} = 250$ GeV, and another at small $m_{\chi} = 25,35$ GeV; both regions correspond to a small $\sigma_p^{SI}$. 
The over-coverage observed in the first region is a result of the high-mass degeneracy (for $m_\chi \gg m_N$, ${dR}/{dE_R}$ depends only on $\sigma_p^{SI}/(\mu_p^2 m_\chi)$; refer to Sec.~\ref{DDtheory}). The importance of this effect decreases with increasing cross-section because the slope of the energy spectrum is better resolved with more events, and hence is more sensitive to slight changes in $v_\text{min}$.  The high-mass degeneracy leads to a 1D profile likelihood that can no longer be well approximated by a Gaussian, such that the test statistic $\lambda(m_{\chi})$ defined in Eq.~\eqref{eq:teststatistics} starts to deviate from a chi-square distribution. The difference between the histogram of $\lambda(m_{\chi})$ values from the mock data and the chi-square distribution with 1 degree of of freedom (as predicted by Wilks' theorem) is shown in Fig. \ref{Fig:Wilks} for a high-mass benchmark point suffering from over-coverage ($m_{\chi} = 250$ GeV, $\sigma_p^{SI} = 2.51 \cdot 10^{-10}$ pb; see left-hand side of Fig.~\ref{1Dm}).  For comparison, we also show the same quantity for a benchmark point where the agreement with the predicted chi-square distribution is much better ($m_{\chi} = 50$ GeV, $\sigma_p^{SI} = 10^{-8}$ pb), and whose coverage is exact to within errors. In contrast, for the high-mass point we observe significant discrepancies in the test statistics $\lambda(m_{\chi})$ for values $\lsim 4$, which explains why over-coverage is observed for this benchmark point.

The over-coverage observed at small $m_{\chi}$ and $\sigma_p^{SI}$ is a result of the low number of counts for this benchmark model. Due to the low statistics in the region of parameter space the 1D profile likelihood is no longer well approximated by a Gaussian, hence the asymptotic behaviour of Wilks' theorem is less accurate. The deviation from Wilks' for these benchmark points is qualitatively similar to the red curve in Fig.~\ref{Fig:Wilks}, albeit less extreme.

Coverage improves when the Ge data are added to the analysis, as can be seen in the right panels of Fig.~\ref{1Dm}.  Exact coverage is obtained in most of the parameter space. An exception is observed at $m_{\chi} = 70$ GeV, $\sigma_p^{SI} = 6.31 \times 10^{-10}$ pb for the $95.4\%$ plot, where slight over-coverage is found. Because neighbouring benchmark points are exactly covered, we interpret this as a statistical fluctuation. Both regions of over-coverage identified in the Xe-only case shrink significantly when adding Ge data to the analysis. For both the $68.3\%$ and the $95.4\%$ interval the over-coverage at large $m_{\chi}$ is almost completely eliminated, except at small $\sigma_p^{SI}$ (for the $68.3\%$ interval), for which the total number of expected events is $\mathcal{O}(10)$.  For higher $\sigma_p^{SI}$, over-coverage of high-mass benchmark models is reduced since the likelihood is tighter for a combined analysis of Xe+Ge.  The remaining over-coverage of the $95.4\%$ interval at $m_{\chi} = 250$ GeV, $\sigma_p^{SI} = 1.58 \times 10^{-9}$ pb corresponds to a value of $97.5\%$, which is just above the border of exact coverage at $97.3\%$. However, at lower masses, especially for the $68.3\%$ contour, over-coverage at very low cross-sections $\sigma_p^{SI} \approx 10^{-10}$ pb is not removed.  In general, we find that the possibility of over-coverage remains as long as WIMP parameters are poorly constrained, which occurs most frequently for benchmark points which imply a low expected number of events.  Both problems are resolved to some extent with the addition of data sets from a second experiment.

\begin{figure*}
  \centering
\includegraphics[width=0.495\linewidth,trim = 10mm 70mm 10mm 70mm]{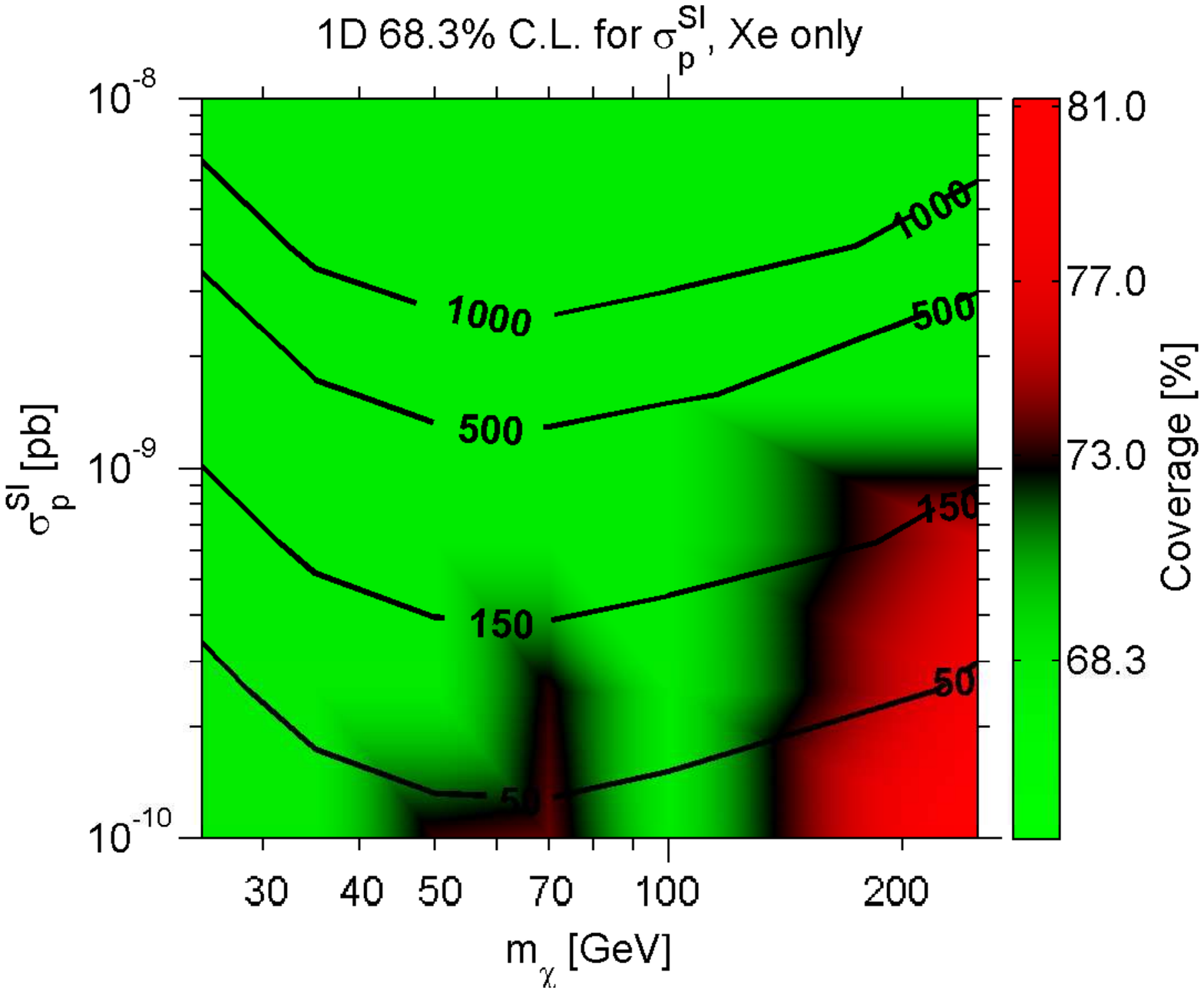} 
\includegraphics[width=0.495\linewidth,trim = 10mm 70mm 10mm 70mm]{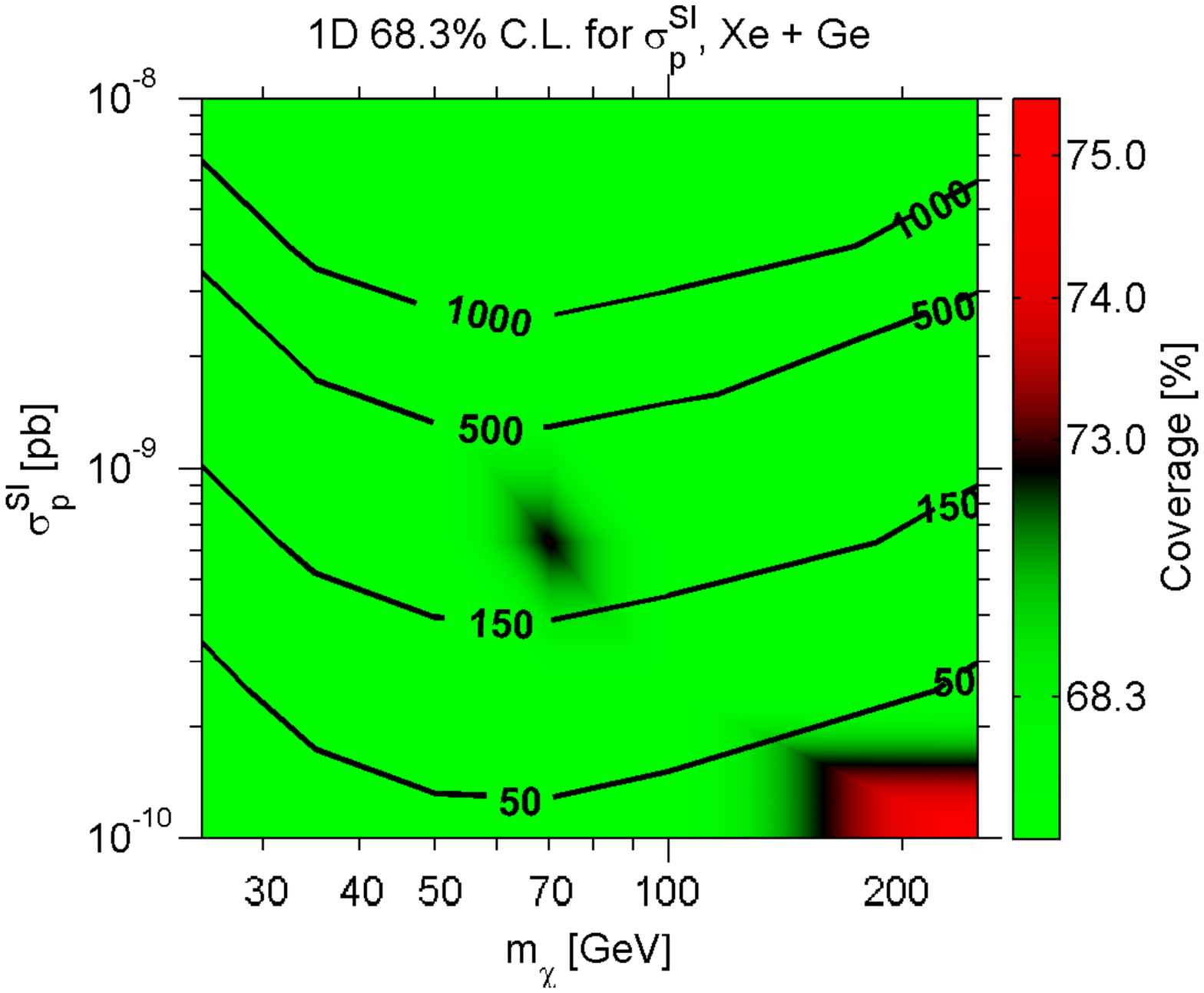} \\
\includegraphics[width=0.495\linewidth,trim = 10mm 70mm 10mm 70mm]{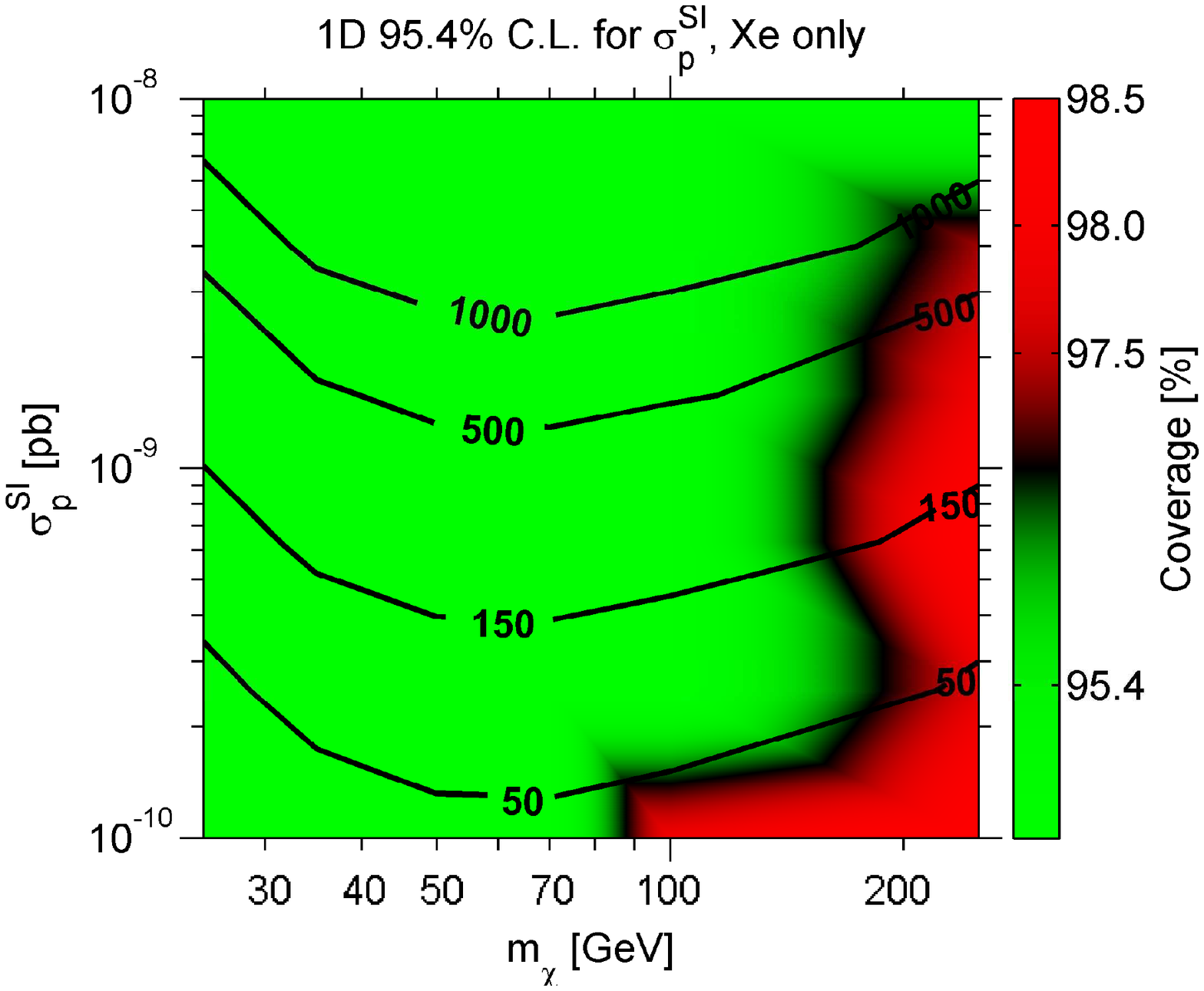} 
\includegraphics[width=0.495\linewidth,trim = 10mm 70mm 10mm 70mm]{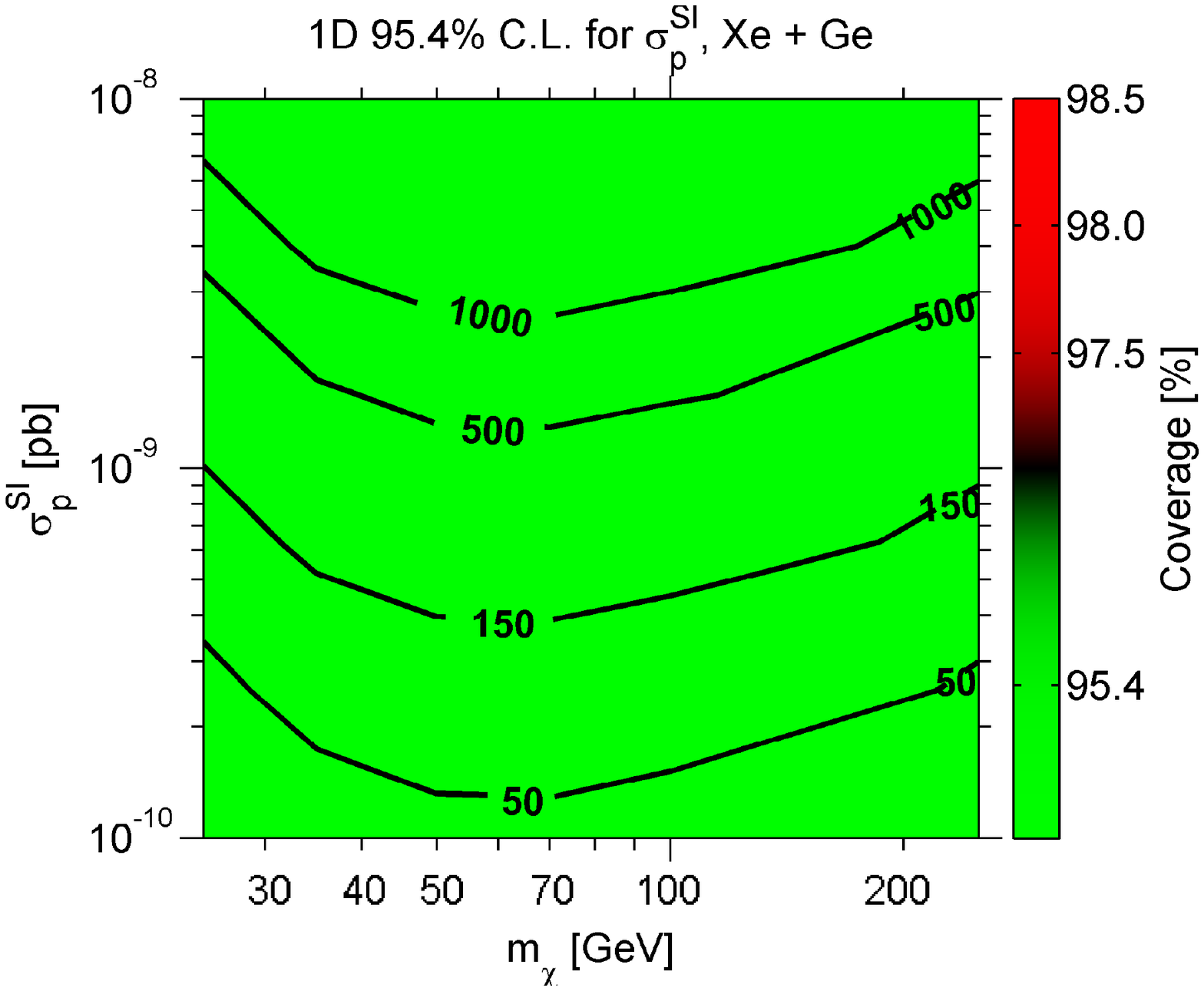} 
\caption{As in Fig.~\ref{1Dm}, but for the 1D confidence intervals for $\sigma_p^{SI}$. A significant improvement in the coverage when combining Xe+Ge is apparent.} \label{1Ds}
\end{figure*}

We display the results of our coverage analysis for the 1D $68.3\%$ and $95.4\%$ confidence intervals for $\sigma_p^{SI}$ in Fig.~\ref{1Ds}. The left-hand plot shows the results for a Xe target, the right-hand plot shows the results for combined Xe+Ge data. In the case in which we consider the Xe data alone, most of the parameter space corresponds to exact coverage, but for both the $1\sigma$ and the $2\sigma$ intervals a large region at high masses $m_{\chi} = 250$ GeV is over-covered. For the $95.4\%$ interval this region is spread over almost the entire cross-section range, and extends to $m_{\chi} = 100$ GeV at low cross-sections. For the $68.3\%$ interval a small region of over-coverage is found at intermediate WIMP masses $m_{\chi} = 50,70$ GeV and low $\sigma_p^{SI}$. For the $95.4\%$ contour the corresponding benchmark points systematically show a coverage percentage at least $1\%$ above the exact value of $95.4\%$. 

The over-coverage at large $\sigma_p^{SI}$ is a result of the high-mass degeneracy, analogously to what has been explained above for the mass.  The over-coverage at intermediate WIMP masses can be explained using Fig.~\ref{bm2fig}. Good reconstructions yield one dimensional profile likelihood functions that are approximately Gaussian, and thus lead to exact coverage. For bad reconstructions, the likelihood is spread over a larger range and thus the statement that $\sigma_p^{SI}$ is over-covered for intermediate WIMP masses is a statement about the ratio of good to bad parameter fits. Due to low statistics resulting from the low number of counts the 1D profile likelihood function can no longer be well approximated by a chi-square distribution, Wilks' theorem becomes less accurate and over-coverage is observed. On the other hand, the over-coverage around 50 GeV WIMPs is not very significant, being close in magnitude to the numerical uncertainty of our coverage values, and therefore could be interpreted as a statistical fluke. 

\begin{table*}[htp]
\centering
\fontsize{9}{9}\selectfont
\begin{tabular}{|c|c|c|c|c|c|c|}
\hline
\hline
\multirow{2}{*}{$m_{\chi}$ [GeV]} & \multirow{2}{*}{$\sigma_p^{SI}$ [pb]} & \multirow{2}{*}{$N_R$} & \multicolumn{4}{|c|}{Coverage [$\%$]} \\\cline{4-7}
&&& 1D $68.3\%$ $m_{\chi}$ & 1D $95.4\%$ $m_{\chi}$ & 1D $68.3\%$ $\sigma_p^{SI}$ & 1D $95.4\%$ $\sigma_p^{SI}$ \\
\hline
$35$ & $10^{-10}$ & 29 & 73.3 (75.4) & 96.1 (96.3) & 69.2 (68.7) & 96.9 (95.5) \\
$50$ & $10^{-10}$ & 38 & 68.3 (73.5) & 95.7 (96.3) & 73.3 (71.2) & 96.9 (96.8) \\
$100$ & $1.58 \times 10^{-9}$ & 527 & 70.3 (69.2) & 96.0 (95.3) & 68.9 (68.4) & 94.9 (95.6) \\
$250$ & $10^{-8}$ & 1671 & 68.0 (66.7) & 95.9 (94.9) & 69.2 (67.6) & 95.7 (95.2) \\
\hline
\end{tabular}
\caption{Results of the coverage analysis of the 1D confidence intervals for four selected benchmark points. Results for the Xe data alone are given, as well as for the combined analysis of Xe+Ge (in parenthesis).} \label{covtable}
\end{table*}

As with the WIMP mass, coverage improves with the addition of data from a Ge target (right plots in Fig.~\ref{1Ds}). For the $68.3\%$ contour the over-covered region at intermediate $m_{\chi} = 50,70$ GeV vanishes completely and is now exactly covered (apart from what can again be interpreted as a statistically non-significant fluctuation around 70 GeV, which appears as a  `flare' pattern in the figure). The over-covered region at high WIMP masses $m_{\chi} = 250$ GeV shrinks significantly, but is difficult to eliminate at low cross-sections $\sigma_p^{SI} = 10^{-10}$ pb, as discussed above. 
The improvement in the coverage is even greater for the $2\sigma$ contour. For a combined analysis of data from Xe+Ge the over-coverage observed for the Xe target completely vanishes; the entire parameter space is exactly covered. The coverage results for a selected subset of benchmark points are shown in Table~\ref{covtable}.

Overall, our coverage analysis concludes that the approximate confidence intervals for the studied benchmark points either cover exactly or over-cover the true values of the parameters -- i.e., they are conservative. The two most important effects at play are the large mass degeneracy, and strong statistical fluctuations that are important even for a relatively large numbers of expected counts ($\sim 100$). We have shown that addition of data from a second target such as Ge leads to significant improvement on both fronts. We point out that the observed over-coverage can in principle be remedied using methods such as Feldman-Cousins to build confidence intervals with guaranteed exact coverage.

We have also investigated coverage properties of the credible intervals obtained from the Bayesian posterior. For well-reconstructed benchmark points, credible intervals are numerically identical to confidence intervals, since we have taken flat priors on our WIMP parameters of interest, so their coverage properties are the same. However, for badly reconstructed points (i.e., lying on the high-mass degeneracy line) the posterior is cut off at large masses and cross-sections by the prior range. This means that the ensuing 1D marginal posterior and thus also the credible intervals become a function of the prior range adopted for the mass and cross section, which is clearly unsatisfactory (this effect has also been pointed out in another context by Ref.~\cite{Arina:2011si}). As a consequence, the coverage of Bayesian credible intervals exhibits broadly the same trends as highlighted above for the frequentist intervals, but also shows a tendency towards under-coverage in some regions. As those results are however sensitive to the choice of prior range, we do not present coverage results for Bayesian credible intervals in this work -- a thorough exploration of this issue would require a study of how such properties change as a function of the prior ranges chosen. We emphasize however that the prior ranges have no impact on our results for the frequentist confidence intervals. 

\subsection{Accuracy and precision of parameter reconstruction}

\begin{figure*} 
  \centering
\includegraphics[width=0.495\linewidth,trim = 10mm 75mm 10mm 75mm]{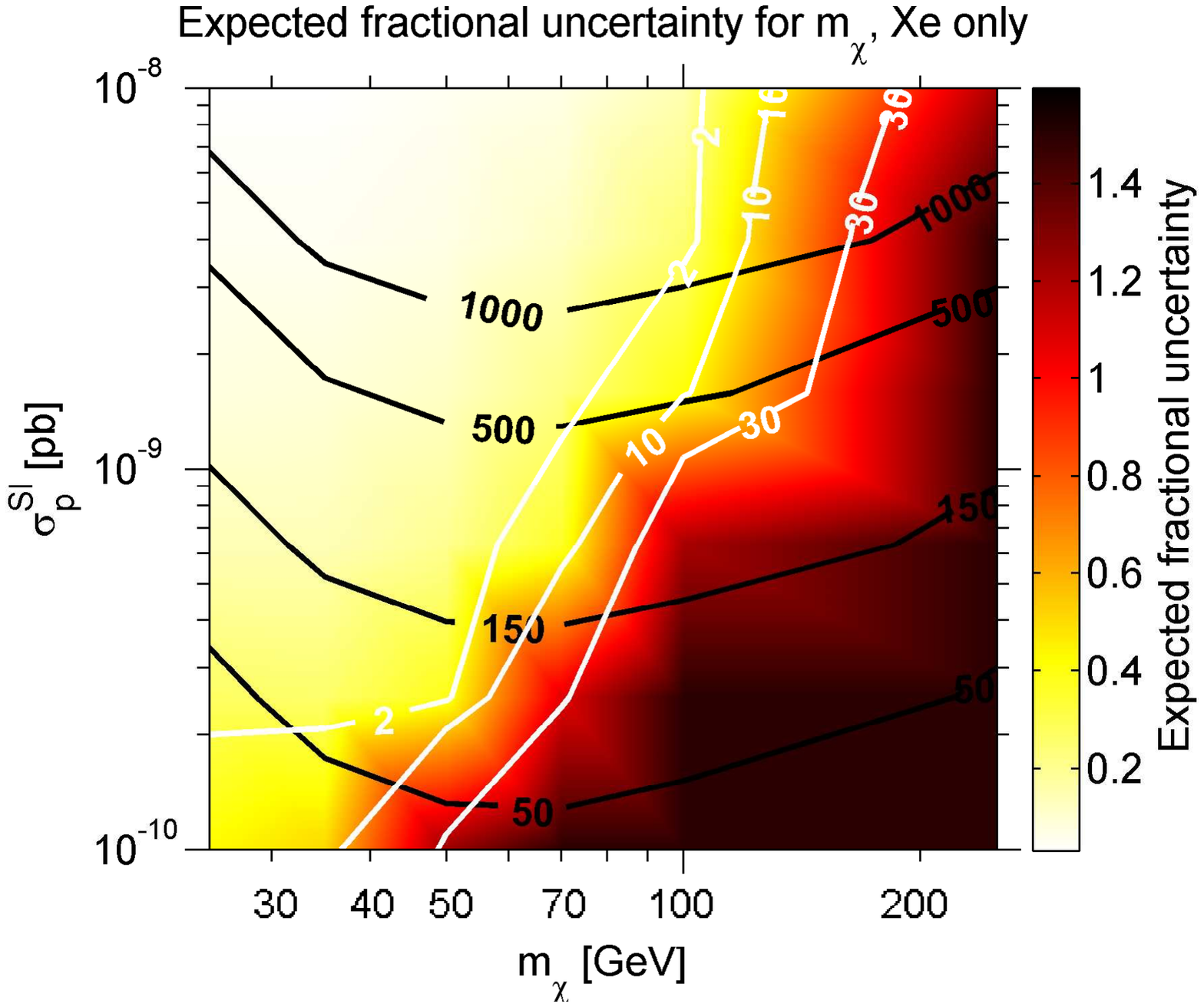} \hfill
\includegraphics[width=0.495\linewidth,trim = 10mm 75mm 10mm 75mm]{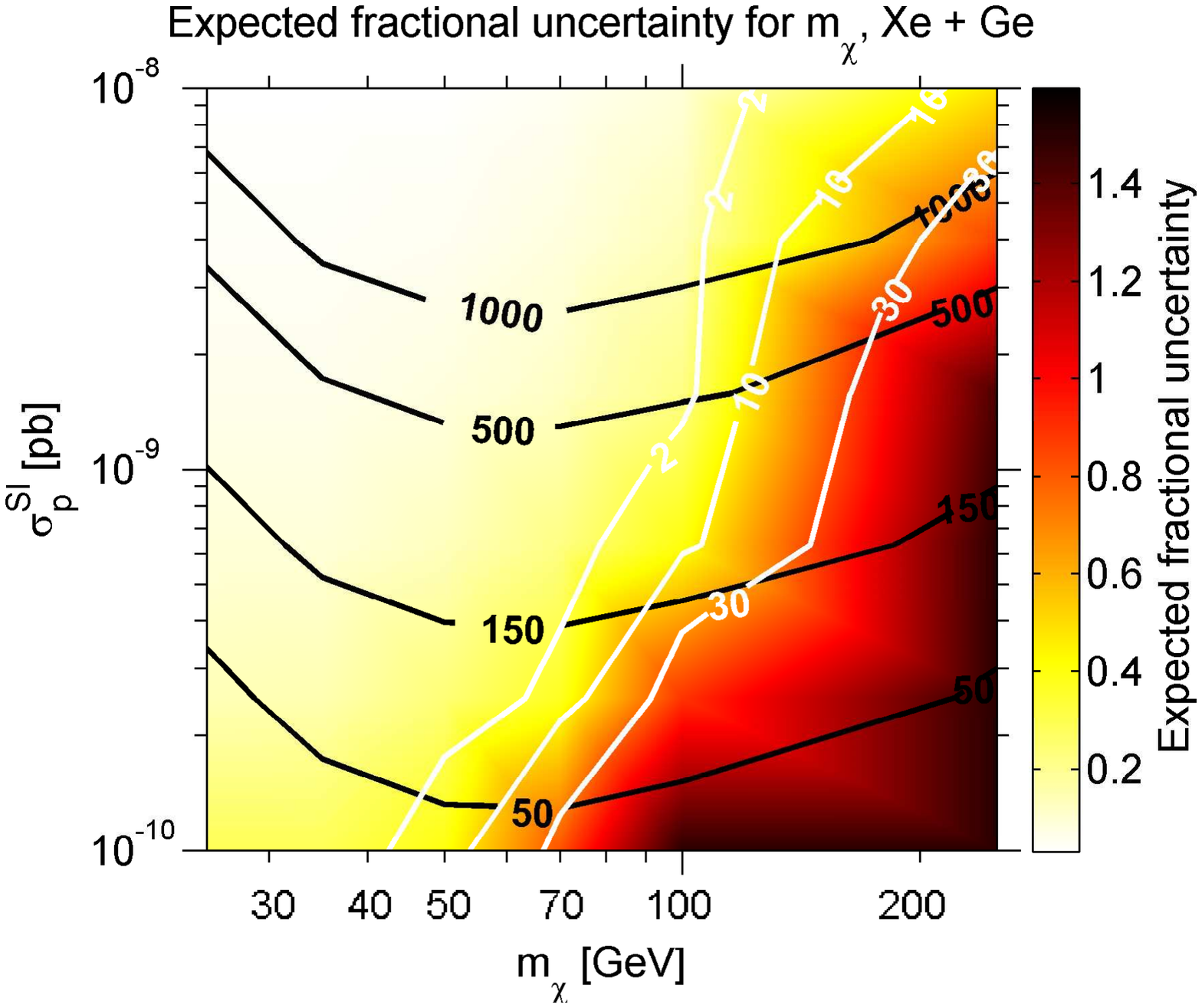} \hfill
\caption{Expected fractional uncertainty (e.f.u.) for the WIMP mass in the $m_{\chi} - \sigma_p^{SI}$ plane, for a Xe (Xe+Ge) target in the left (right) plot, quantifying the precision of the mass reconstruction (low e.f.u.\ corresponding to better precision). Isocontours of the expected number of counts in the Xe experiment are given in black; isocontours of the percentage of ``bad'' reconstruction (f.u. $> 0.75$) are shown in white.} \label{efaplot}
\end{figure*}

We now consider the question of the accuracy and precision of the parameter reconstruction.  We start by investigating the expected fractional uncertainty (e.f.u.) for $m_{\chi}$, introduced in section \ref{statstests}. The e.f.u.\ quantifies the average fractional standard deviation of the reconstructed WIMP mass value and thus is a measure of the precision of the reconstruction.  We show the e.f.u.\ in the $m_{\chi} - \sigma_p^{SI}$ plane in Fig.~\ref{efaplot} (notice that the upper limit of the colorbar is set to e.f.u.$ = 1.5$ for display purposes, but this limit is surpassed in many cases). Isocontours of the expected number of counts in a Xe target are shown in black. Isocontours of the number of ``bad'' cases (i.e., with an f.u. $>0.75$) are shown in white. Considering the number of ``bad" cases is very important, since this number quantifies the probability that, for a given WIMP benchmark point (that may lead to a reasonably small {\it average} uncertainty on $m_{\chi}$), the experiment results in a data set that leaves the WIMP mass essentially unconstrained.

High-mass benchmark points lead to a likelihood function with a long tail in the $m_{\chi} - \sigma_p^{SI}$ plane, and thus are expected to have a very high e.f.u.. We are most interested in the region where the transition from good to poor performance takes place.  

We will first discuss the e.f.u.\ results from Xe data only. As a general pattern, the larger $m_{\chi}$ and the smaller $\sigma_p^{SI}$, the larger the e.f.u.\ value for the benchmark point. We will discuss the e.f.u.\ results at high ($\sigma_p^{SI} = 10^{-8}$ pb), intermediate ($\sigma_p^{SI} = 10^{-9}$ pb) and low ($\sigma_p^{SI} = 10^{-10}$ pb) cross-sections. 

At high ($\sigma_p^{SI} = 10^{-8}$ pb) cross-sections, most benchmark masses lead to a small e.f.u., and thus a small uncertainty in the reconstructed WIMP mass. The e.f.u.\ does not exceed $0.15$ for $m_{\chi} \leq 100$ GeV and is significantly smaller for small $m_{\chi} = 25,35$ GeV (e.f.u.\ $= {0.03}$). The fraction of bad reconstructions is $< 1\%$. However, even for this large cross-section and the resulting large number of events, $N_R = 1671$, the high-mass benchmark point $m_{\chi} = 250$ GeV leads to an e.f.u.\ $> 1.00$. Such a large e.f.u.\ means that the WIMP mass is left essentially unconstrained by the data, and the confidence levels inhabit the region of degeneracy at high masses and cross-sections.

For intermediate benchmark cross-sections ($\sigma_p^{SI} = 10^{-9}$ pb), the overall precision is quite good. For benchmark masses $m_{\chi} \leq 70$ GeV the e.f.u.\ is $<0.30$ and the WIMP mass is well constrained. This is also reflected in the number of bad reconstructions: for $m_{\chi} \leq 50$ GeV this number is $<1\%$; for $m_{\chi} = 70$ GeV only a couple of bad cases occur for 100 reconstructions. At higher $m_{\chi}$ the e.f.u.\ increases rapidly. For example, at $m_{\chi} = 100$ GeV the e.f.u.\ increases from $0.41$ to $1.21$ when decreasing the cross-section from $\sigma_p^{SI} = 1.58 \times 10^{-9}$ (corresponding to $N = 527$ events) to $\sigma_p^{SI} = 6.31 \times 10^{-10}$ (corresponding to $N = 210$ events). Therefore, at $\sigma_p^{SI} = 10^{-9}$ this benchmark point lies on the borderline between good and bad performance of the reconstruction. At cross-sections $\sigma_p^{SI} \leq 10^{-9}$ and high WIMP masses ($m_{\chi} \geq 100$ GeV), the e.f.u.\ is systematically $>$0.75 (sometimes $\gg$0.75), meaning that the WIMP mass becomes essentially unconstrained in $20\%$ or more of the reconstructions. This is to be expected, due to the $m_{\chi} - \sigma_p^{SI}$ degeneracy that occurs at high masses. However, it is interesting to see how pronounced this effect is even at a relatively small mass ($m_{\chi} \approx 100$ GeV). 

The situation deteriorates significantly for $\sigma_p^{SI} = 10^{-10}$ pb, leading to a small number of counts [$\mathcal{O}(10)$] for all $m_{\chi}$. This is reflected in the e.f.u., which is of order $\sim$0.50 for small $m_{\chi} = 25,35$ GeV.  This corresponds to weak constraints on the WIMP mass, and leads to an average uncertainty of more than $100\%$ for $m_{\chi} \geq 50$ GeV. Similarly, while for small WIMP masses just above $5\%$ of all reconstructions are bad, this number is significantly higher for high-mass WIMP models. Even for an intermediate $m_{\chi} = 50$ GeV, $\sim$30\% of reconstructions are bad. We emphasize once more that this is due to statistical fluctuations in the realization of the energy spectrum, and therefore an unavoidable effect.

As expected, the e.f.u.\ improves considerably with the addition of data from a Ge target.  For fixed cross section, the 30\% bad reconstruction isocontour shifts to higher mass values by $\sim 50\%$ with respect to the reconstruction with Xe data alone.  Because the e.f.u.\ is correlated with the percentage of poor reconstructions, we also see that it decreases dramatically at fixed WIMP parameters (often by $> 50\%$) with the inclusion of the Ge data.

\begin{figure*} 
  \centering
\includegraphics[width=0.9\linewidth,trim = 10mm 90mm 10mm 90mm]{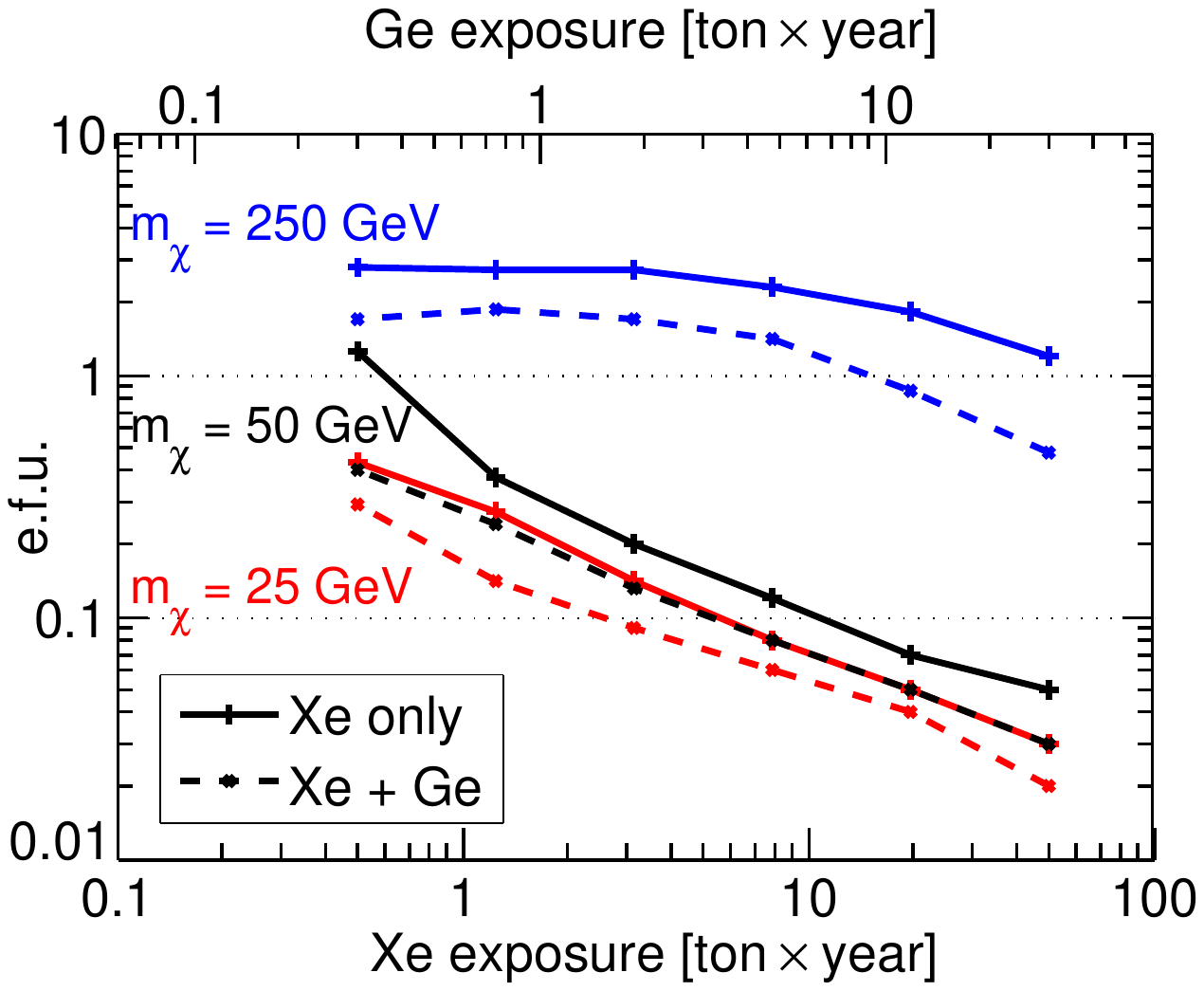}
\caption{Expected fractional uncertainty (e.f.u.) on the WIMP mass as a function of exposure for a Xenon experiment (bottom axis) and a Germanium experiment (top axis) required to achieve this e.f.u.\ for a WIMP with cross-section $\sigma_p^{SI} = 10^{-9}$, for three different benchmark masses $m_{\chi} = 25$ GeV (red), $m_{\chi} = 50$ GeV (black) and $m_{\chi} = 250$ GeV (blue). Solid lines correspond to e.f.u.\ results for Xe only, dashed lines correspond to e.f.u.\ results for a Xe + Ge target.} \label{efu_counts}
\end{figure*}

Fig.~\ref{efu_counts} shows the value of the e.f.u.\ as a function of the exposure $\epsilon$ for a WIMP with cross-section $\sigma_p^{SI} = 10^{-9}$ pb and for three different benchmark masses. Solid lines correspond to the e.f.u. from a Xe target only, dashed lines show results for combining data from a Xe and a Ge experiment. For the Xe only case, for massive WIMPs ($m_{\chi}  = 250$ GeV), the expected fractional uncertainty is always greater than unity, as a consequence of the degeneracy. For intermediate ($m_{\chi}  = 50$ GeV) and small mass WIMPs ($m_{\chi}  = 25$ GeV), the e.f.u.\ drops sharply with increasing exposure. In particular, it is still of order $\sim 30-40\%$ for an exposure of 1 ton$\times$year, and it is reduced to less than 10\% for a Xe experiment with exposure 10 ton$\times$year. 
When combining Xe + Ge data the situation improves for all benchmark masses. For massive WIMPs ($m_{\chi}  = 250$ GeV) an e.f.u. smaller than unity can be achieved for a Xe experiment with exposure $\sim20$ ton$\times$year and a Ge experiment with exposure $\sim10$ ton$\times$year. For larger exposures the  e.f.u. further decreases. For both intermediate ($m_{\chi}  = 50$ GeV) and small ($m_{\chi}  = 25$ GeV) WIMP masses the e.f.u. for Xe + Ge is significantly smaller than in the Xe only case. The e.f.u. strongly decreases as the exposures of the Xe and Ge targets are increased. In particular, for an intermediate (low) mass WIMP an expected fractional uncertainty of less than 10\% can be achieved for a 3 (1.5) ton$\times$year exposure for Ge and a 5 (3) ton$\times$year exposure for Xe.  These trends are qualitatively consistent with those found by Refs. \cite{green2008,drees2008}.

However, we caution that the e.f.u. will be higher in reality for a fixed exposure and benchmark point, because of astrophysical and nuclear-physics uncertainties.

\begin{figure*} 
  \centering
\includegraphics[width=0.495\linewidth,trim = 10mm 75mm 10mm 75mm]{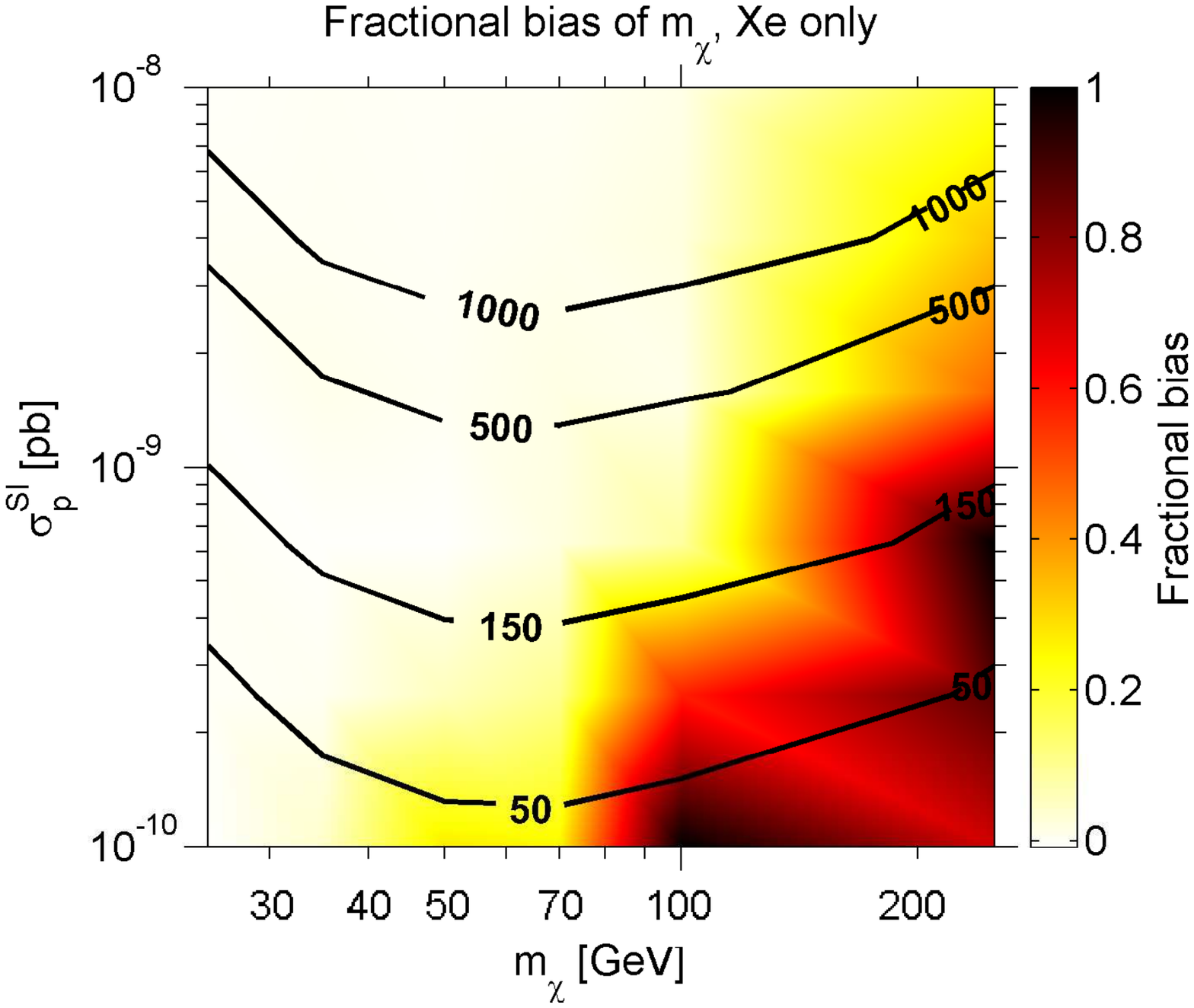} \hfill
\includegraphics[width=0.495\linewidth,trim = 10mm 75mm 10mm 75mm]{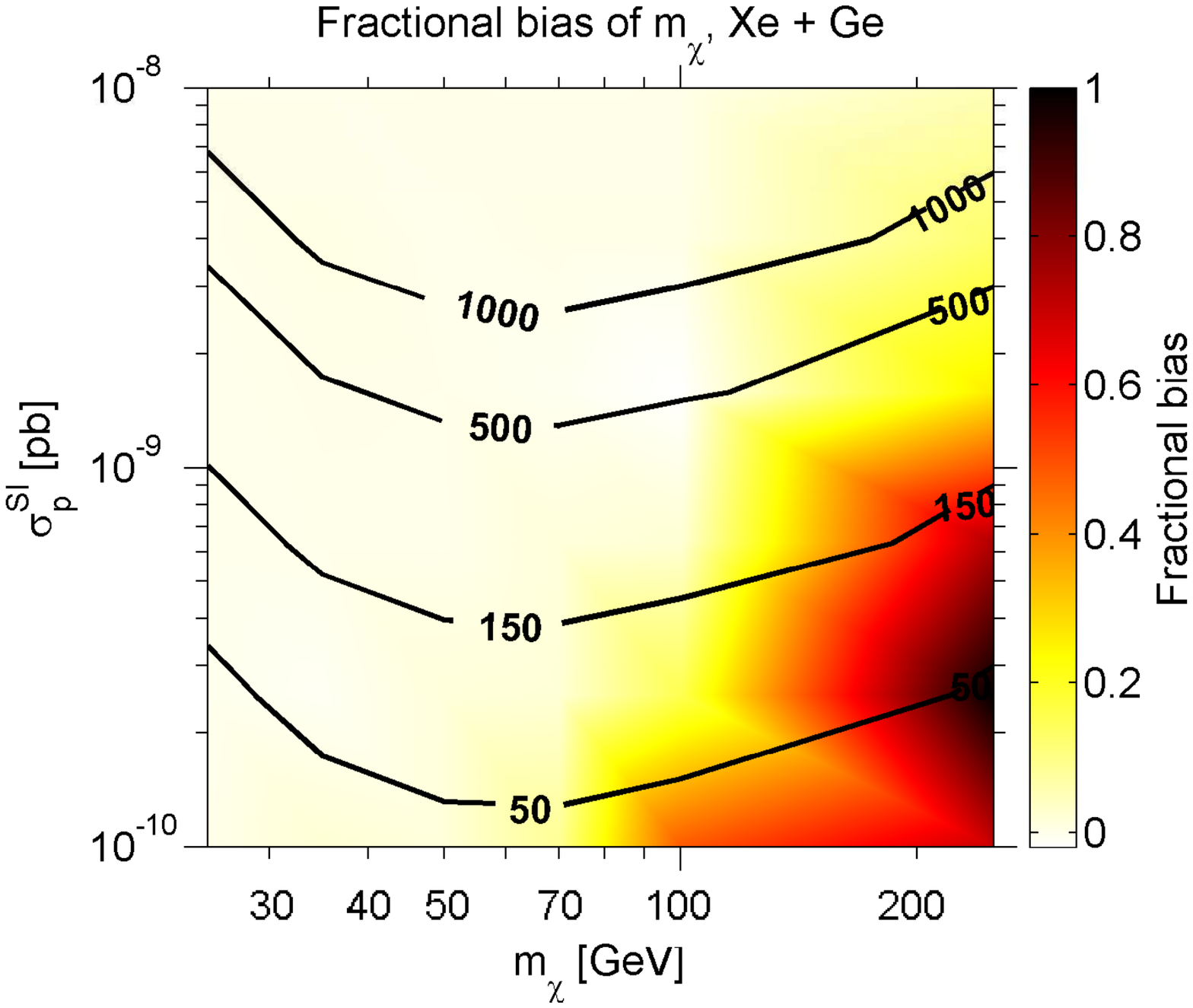} \hfill
\caption{As in Fig.\ref{efaplot}, but for the fractional bias of the WIMP mass, i.e. the bias of the WIMP mass relative to the benchmark mass (notice that almost no negative bias is observed).} \label{bias}
\end{figure*}

The fractional mass bias in the $m_{\chi} - \sigma_p^{SI}$ plane for a Xe target (Xe and Ge target) is displayed on the left (right) of Fig.~\ref{bias}. Almost no negative bias in the mass is observed. If a bias exists, it typically goes in the direction of a larger $m_{\chi}$ than the true value, as a consequence of the high mass-cross section degeneracy. In fact, the distribution of reconstructions that reach up onto the degeneracy curve explains the features of Fig.~\ref{bias}.  In comparing Figs. \ref{efaplot} and \ref{bias}, we find that the curve for e.f.u.\ = 0.8 corresponds closely to the curve of bias = 0.2.  When a large fraction of reconstructions are bad, both the e.f.u.\ and bias increase because the high mass-cross section curve becomes populated with high-likelihood fits.  The extension of the confidence levels to this region of the parameter space means that the best-fit mass is typically higher than the true mass, so that both the uncertainty in the mass and its bias become large.

The performance of the statistical reconstruction (as quantified by the e.f.u., the number of bad cases and the fractional bias in the WIMP mass) is summarised for four benchmark points in Table~\ref{statstable}.
\begin{table*}[htp]
\centering
\fontsize{9}{9}\selectfont
\begin{tabular}{|c|c|c|c|c|c|}
\hline
\hline
 $m_{\chi}$ [GeV] &  $\sigma_p^{SI} [pb]$ & $N_R$ & e.f.u.\ & \# bad cases & fractional bias for $m_{\chi}$ \\
\hline
$35$ & $10^{-10}$ & 29 & 0.51 (0.29) & 7 (0) & 0.042 (0.023) \\
$50$ & $10^{-10}$ & 38 & 1.24 (0.40) & 32 (4) & 0.272 (0.017) \\
$100$ & $1.58 \times 10^{-9}$ & 527 & 0.41 (0.22) & 9 (0) & 0.014 (-0.020) \\
$250$ & $10^{-8}$ & 1671 & 1.20 (0.48) & 51 (13) & 0.205 (0.052) \\
\hline
\end{tabular}
\caption{Summary of the performance of the statistical reconstruction four selected WIMP benchmark models. The benchmark (true) mass and cross-section and the corresponding number of counts for the Xe experiment are shown. We give the expected fractional uncertainty, the number of ``bad'' (f.u. $> 0.75$) cases and the fractional bias in $m_{\chi}$ for the Xe data alone and for the combined analysis of Xe+Ge (in parenthesis).}\label{statstable}
\end{table*}

\subsection{Comparison with other coverage studies}

We have focused on reconstructing phenomenological WIMP-related variables (mass, spin-independent cross section) rather than theoretical parameters in specific theories for WIMP physics.  Perhaps not surprisingly, our results differ from recent studies of the coverage properties of parameters of specific supersymmetric models from particle-physics experiments, including direct-detection data~\cite{Akrami:2010cz,Bridges:2010de}.   Ref.~\cite{Bridges:2010de} 
found that supersymmetric parameters were consistently over-covered  when attempting to reconstruct the `SU3' benchmark point with mock ATLAS data on sparticle masses and mass splittings.  In contrast, consistent (and sometimes drastic) under-coverage was observed \cite{Akrami:2010cz} for two different benchmark points reconstructed using mock ton-scale direct detection data.  

Here, we observed exact coverage in a large portion of the phenomenological parameter space we investigated. Unlike in supersymmetric analyses, the parameter space considered here does not include complicated theoretical boundaries where the likelihood function is not defined.  Substantial over-coverage is therefore not expected in our results for cases with reasonable statistics (i.e.~where Wilks' Theorem does not break down simply due to low-number statistics).  Furthermore, the relationship between parameters of interest (here, WIMP mass and cross-section) and observables (i.e., counts) is far simpler here than when one works with fundamental supersymmetric parameters (which are connected to observables via complex, non-linear Renormalization Group Equations that make the likelihood function highly non-Gaussian in the parameters). Therefore, sampling issues that might plague supersymmetric parameter spaces and lead to under-coverage are not observed in our setup.

Taking the results of all three studies together, we conclude that coverage properties are good when the scanning is done over a set of parameters that have a simple mapping to the observables (as was seen in \cite{Bridges:2010de}).  As the observables on which a (typically approximately Gaussian) likelihood function is defined become a highly complicated function (i.e.~via highly non-linear transformations) of the parameters of interest, the coverage becomes less exact, and a detailed numerical investigation is required to establish the coverage properties. The upshot of this for dark matter searches is that simple model-independent analyses using phenomenological particle-physics parameters for WIMPs can generally be expected to have good coverage, but the mapping onto specific model spaces will typically not retain this property.

\section{Conclusions}
\label{sec:summary}

We have studied the statistical properties of approximate confidence intervals on WIMP parameters, using mock data from future ton-scale direct detection experiments.  We have focused in particular on the effect of unavoidable statistical fluctuations in the data. Contrary to what has been observed in GUT-scale SUSY parameterisations, we see that coverage for phenomenological WIMP parameters (mass, cross-section) is generally quite good. We have observed a small amount of over-coverage for certain benchmark points, i.e.~the constructed confidence intervals are conservative. We have traced this over-coverage back to either statistical fluctuations, which become most important for benchmark points leading to a low expected number of counts, or to the degeneracy between the WIMP mass and cross-section, that occurs at large WIMP masses in the likelihood function. In both cases the profile likelihood is not well approximated by a Gaussian, such that Wilks' theorem no longer accurately described the behaviour of the test statistics $\lambda(m_\chi)$ and $\lambda(\sigma_\mathrm{SI})$.  This problem is much less severe than in the SUSY case; in general, it appears that the less complicated and nonlinear a function the likelihood is of the underlying parameter space, the better the coverage properties.  Finally, we remind the reader that coverage issues can in principle be resolved altogether by constructing intervals that have exact coverage, e.g.~by using the Feldman-Cousins method.

We have found that the statistical bias and expected fractional uncertainty of the reconstructed WIMP mass and cross-section are more serious problems, which cannot be resolved by employing a different method of constructing confidence intervals. The parameter reconstruction can be ruined by statistical fluctuations that flatten the observed energy recoil spectrum with respect to the true underlying model, leading to an essentially unconstrained likelihood function, so that only a lower limit can be placed on the WIMP mass and cross-section. This was found to be important even at intermediate WIMP masses and cross-sections. Therefore, even for benchmark models leading to a relative large expected number of counts ($\gtrsim \mathcal{O}(100)$), statistical fluctuations can result in a strong bias (i.e. low accuracy) and a low precision of the reconstruction of the WIMP parameters.

We have shown that a combination of data sets from two independent experiments with different target materials can significantly improve the coverage properties, reduce the bias and increase the accuracy and precision of the reconstruction. Furthermore, we have shown that the precision of the reconstruction can be improved considerably if the exposure of the experiment(s) is increased.

Our investigation has assumed negligible backgrounds and fixed important sources of uncertainties, such as astrophysical quantities describing the local dark matter distribution. Our modelling of the experimental likelihood was correspondingly simplified. Therefore, the large bias and low precision of the reconstructed parameters discovered for a number of benchmark models is a fundamental result of statistical fluctuations in the realisation of the energy spectrum. We expect that including the energy resolution, non-negligible backgrounds and astrophysical uncertainties in the analysis would further degrade the performance of the reconstruction. 

\section*{Acknowledgments}
We would like to thank Miguel Pato, Alastair Currie and Henrique Araujo for useful discussions. C.S. and R.T. thank the University of Zurich for hospitality. R.T.~would like to thank Kavli Institute for Theoretical Physics at the UCSB for hospitality. C.S. would like to thank the McGill High Energy Theory Group for hospitality. C.S. is partially supported by a scholarship of the ``Studienstiftung des deutschen Volkes''. This research was supported in part by the National Science Foundation under Grant No. NSF PHY05-51164. A.H.G.P.~is supported by a Gary McCue Fellowship through the Center for Cosmology at UC Irvine and by NASA Grant No. NNX09AD09G.  P.S. is supported by the Lorne Trottier Chair in Astrophysics and an Institute for Particle Physics Theory Fellowship. G.B. is supported by a European Research Council Starting Grant, under grant agreement No. 277591.

\bibliography{references}

\end{document}